\documentclass[a4paper,fleqn,usenatbib]{mnras}

\usepackage[T1]{fontenc}
\usepackage{ae,aecompl}

\usepackage{graphicx}	% Including figure files
\usepackage{amsmath}	% Advanced maths commands
\usepackage{amssymb}	% Extra maths symbols
\usepackage[usenames,dvipsnames]{color}

\newcommand{\cena}{NGC\,5128}

\title[The Survey of Centaurus A's Baryonic Structures]{The Survey of Centaurus\,A's Baryonic Structures (SCABS). I. Survey Description and Initial Source Catalogues}

\author[M. A. Taylor et al.]{Matthew A. Taylor,$^{1,2}$\thanks{Visiting astronomer, Cerro Tololo Inter-American Observatory, National Optical Astronomy Observatory, which is operated by the Association of Universities for Research in Astronomy (AURA) under a cooperative agreement with the National Science Foundation. E-mail: mtaylor@astro.puc.cl (MAT)}
Roberto P. Mu\~noz,$^{1}$
Thomas H. Puzia,$^{1}$
Steffen Mieske,$^{2}$
\newauthor{Paul Eigenthaler,$^{1}$
and Mia Sauda Bovill$^{3}$}
\\
$^{1}$Institute of Astrophysics, Pontificia Universidad Cat\'olica de Chile, Av.~Vicu\~na Mackenna 4860, 7820436 Macul, Santiago, Chile\\
$^{2}$European Southern Observatory, Alonso de Cordova 3107, Vitacura, Santiago, Chile\\
$^{3}$Space Telescope Science Institute, 3700 San Martin Drive, 21218, Baltimore, Maryland, USA
}

\date{Accepted XXX. Received YYY; in original form \today}

\pubyear{2016}

\begin{document}
\label{firstpage}
\pagerange{\pageref{firstpage}--\pageref{lastpage}}
\maketitle

\begin{abstract}
We present new, wide-field, optical ($u'g'r'i'z'$) {\it Dark Energy Camera} observations covering $\sim21\,{\rm deg}^2$ centred on the nearby giant elliptical galaxy \cena\ called ``The Survey of Centaurus A's Baryonic Structures'' (SCABS). The data reduction and analysis procedures are described including initial source detection, photometric and astrometric calibration, image stacking, and point-spread function modelling. We estimate 50 and 90 percent, field-dependent, point-source completeness limits of at least $u'=24.08$ and $23.62$\,mag (AB), $g'=22.67$ and $22.27$\,mag, $r'=22.46$ and $22.00$\,mag, $i'=22.05$ and $21.63$\,mag, and $z'=21.71$ and $21.34$\,mag. Deeper imaging in the $u'$-, $i'$- and $z'$-bands provide the fainter limits for the inner $\sim3\,{\rm deg}^2$ of the survey, and we find very stable photometric sensitivity across the entire field of view. Source catalogues are released in all filters including spatial, photometric, and morphological information for a total of $\sim5\times10^5-1.5\times10^6$ detected sources (filter-dependent). We finish with a brief discussion of potential science applications for the data including, but not limited to, upcoming works by the SCABS team. 
\end{abstract}


%%%%%%%%%%%%%%%%%%%%%%%%%%%%%%%%%%%%%%%%%%%%%%%%%%

%%%%%%%%%%%%%%%%% BODY OF PAPER %%%%%%%%%%%%%%%%%%

\section{Introduction}
Since the turn of the century, the Sloan Digital Sky Survey \citep[SDSS;][]{yor00}, with its thousands of deg$^2$ of sky coverage, has demonstrated the utility of modern CCD-based large-scale imaging campaigns. As a result, a new generation of  wide-field imaging cameras has enabled relatively small research groups to conduct intermediate-scale surveys capable of deeply imaging areas of sky ranging from dozens to 100s of deg$^2$ that were previously accessible to much larger consortia of researchers. These imagers are particularly useful to conduct deep studies of nearby galaxy groups and clusters, which are rapidly revealing the never before accessible faint properties of these systems \citep[e.g.][]{chi09,mcc09,fer12,mer14,mun14,mun15}.

Among the most well-known such structures in the nearby universe is the Centaurus A group, which is dominated by the giant galaxy \cena\ at a distance of $3.8\pm0.1\,{\rm Mpc}$ \citep[][]{har10}. The group is mostly comprised of at least 40 known dwarf galaxies \citep[][]{cot97,van00,kar07,crn14,crn15,tul15}, which with the new wide-field cameras coming online has been recently attracting a new wave of attention. One recent example is the PISCeS survey \citep{san14,crn14,crn15}, which images $\sim11\,{\rm deg}^2$ around \cena\ in the optical $g'$- and $r'$-bands using the optical Megacam imager at the 6.5-m Magellan II Clay telescope \citep{mcleod15}. This survey has recently revealed nine new low-surface brightness ($25.0\lesssim\mu_{r,0}\lesssim27.3\,{\rm mag}; -13\lesssim M_V\lesssim -7.2\,{\rm mag}$) dwarf galaxies within $\sim150$\,kpc of \cena. A yet more ambitious program uses the {\it Dark Energy Camera} \citep[DECam;][]{fla15} to cover $\sim550\,{\rm deg}^2$ encompassing \cena, and the nearby M\,83 complex \citep[][hereafter MJB15/16]{mul15, mul16}. This project has potentially more than doubled the known population of dwarf galaxies around the two giants, identifying at least 57 new candidates as faint as $\mu_{r,0}\approx29\,{\rm mag}\,{\rm arcsec}^{-2}$. While these candidates await confirmation via spectroscopy and/or resolved stellar population studies, the utility of wide-field imagers like DECam in revealing the secrets of these iconic neighbours is clear.

In this contribution we present a new imaging campaign of $\sim\!72\,{\rm deg}^2$ centred on \cena\ using DECam at the 4-m Baade telescope at the Cerro Tololo Interamerican Observatory (CTIO) in Chile. Our program, {\it The Survey of Centaurus A's Baryonic Structures} (SCABS) is complementary to the other two recent large-scale projects. Among the most powerful features of the PISCeS campaign is the ability to conduct resolved stellar population studies, combined with depth that samples the red giant branch at the distance of \cena. On the other hand, while the MJB15/16 program reaches similar depths as PISCeS, they do not benefit from their resolution capabilities, but probe an area $\sim50\times$ greater. Noting the strengths and shortcomings of these two surveys, SCABS addresses complementary science goals by homogeneously imaging \cena\ to $\sim2\times$ the galactocentric radius (out to $R_{\rm gc}\approx300\,{\rm kpc}$) compared to PISCeS, which has so-far been focussed mainly on the NE quadrant of the galaxy. More importantly, while we only sample a fraction of the area covered by MJB15/16, we do so in the five optical $u'$, $g'$, $r'$, $i'$, and $z'$ filters, providing significant spectral energy distribution (SED) leverage and opening the door to a suite of ancillary science goals not possible with $g'$- and $r'$-band imaging alone.

The paper is organized as follows. \S\,\ref{sec:obs} gives an overview of the SCABS observations, including a brief description of DECam's capabilities. \S\,\ref{sec:reduction} focusses on the data reduction, image processing, and photometric measurements carried out on the inner $\sim21\,{\rm deg}^2$ region of SCABS upon which this work is based. We also describe artificial star experiments which were used to asses the photometric quality and depth of our observations. \S\,\ref{sec:disc} describes the public release of our five-band source catalogues, and presents several science cases and applications for the SCABS data. Throughout this work we adopt a distance modulus for \cena\ of $m-M=27.88\pm0.05\,{\rm mag}$, corresponding to a distance of $3.8\pm0.1\,{\rm Mpc}$ \citep{har10}.

%%%%%%%%%%%%%%%%%%%%%%%%%%%%%%
%%%%%%%%%%%%%%%%%%%%%%%%%%%%%%
%%%%%%%%%%%%%%%%%%%%%%%%%%%%%%
\section{Observations}
\label{sec:obs}
The $u'$, $g'$, $r'$, $i'$, and $z'$-band SCABS data were taken during the nights of 2014 April 4--5 (CNTAC\,ID: 2014A-0610; PI: Matthew Taylor), and 2014 August 25--27 (CNTAC\,ID: 2014B-0609; PI: Roberto Mu\~noz) using DECam mounted at the prime focus on the 4\,m Victor Blanco telescope at the Cerro Tololo Inter-American Observatory (CTIO) in Chile. DECam is a wide field imager equipped with the optical {\it u', g', r', i', z'} and Y filters. Working at a focal ratio of $f/2.7$, it is comprised by 62 2\,048$\times$4\,096\,pixel, two-amplifier imaging CCDs, each separated by 201 and 153\,pix ($\sim$53 and 40$''$) gaps along the long and short edges respectively, giving a spatial coverage of $\sim2.9\,{\rm deg}^2$. One chip (N30) shows poor charge transfer efficiency rendering it scientifically useless, and since 30 November 2013 chip S30 stopped working. Additionally the Southern half of S7 malfunctions, and thus DECam is effectively comprised of 59.5 CCDs with a total of $\sim500$\,Mpix at pixel scales between 0.2626\,$''$/pix at the edge of the field-of-view, to 0.2637\,$''$/pix at the centre. Luckily, the positions of the bad chips make it easy to recover the full spatial extent of DECam in image mosaics that employ an appropriate dithering pattern. For more details on the DECam technical specifications we refer the reader to the online documentation\footnote{\url{http://www.ctio.noao.edu/noao/content/DECam-What}}.

\begin{figure*}
	\includegraphics[width=\linewidth]{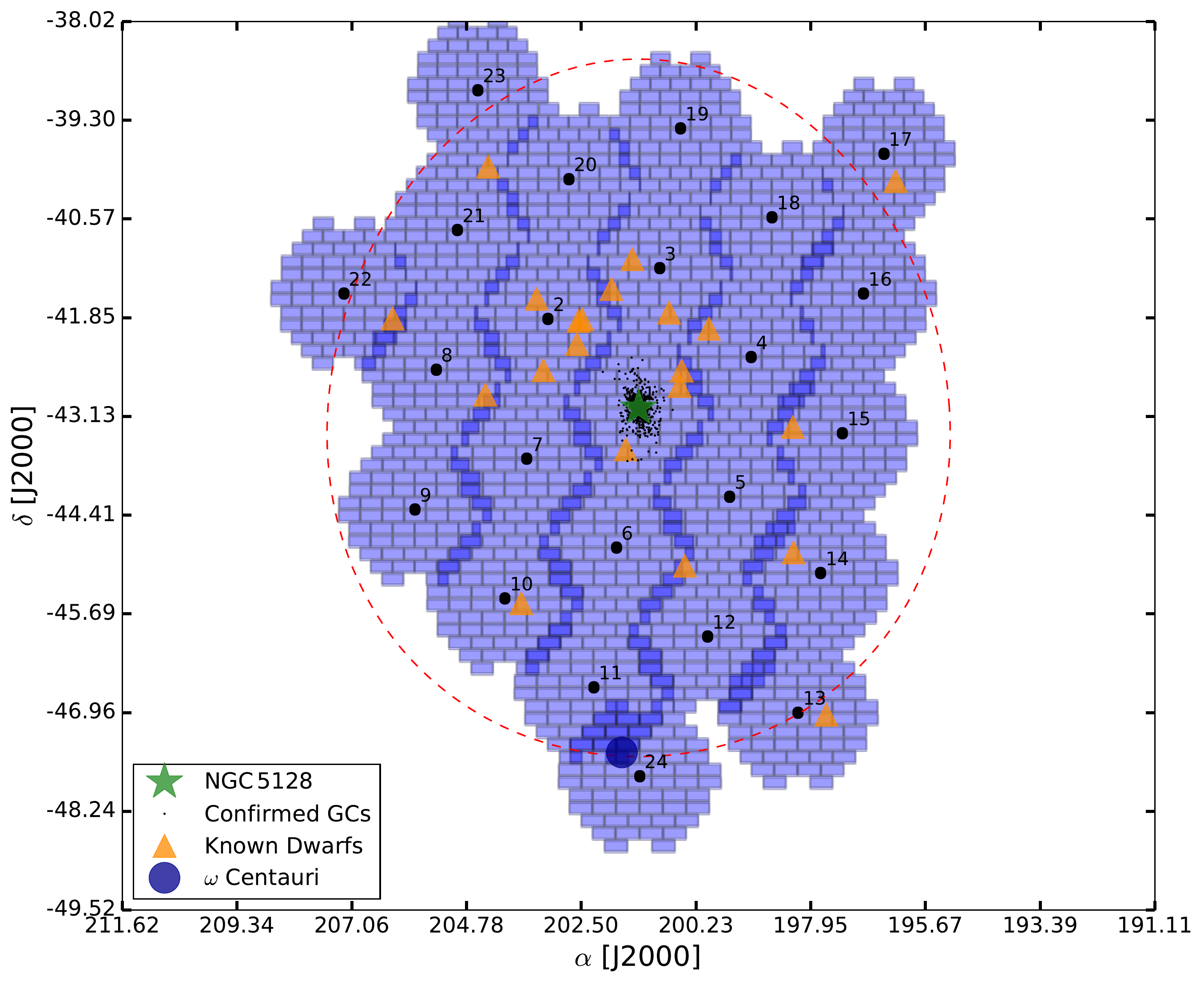}
	\caption{The spatial coverage of the SCABS observations. The position of \cena\ is shown by the green star, while the surrounding cloud of black dots indicates the population of radial velocity confirmed GCs. Orange triangles denote the positions of the previously known dwarf galaxy population within the SCABS footprint, and the position of the galactic GC $\omega$-Centauri is indicated by the maroon circle, which serendipitously falls within the overall SCABS footprint. Different tiles are indicated by the numbers shown, with Tile\,1 centred on \cena\ itself.}
	\label{fig:dither}
\end{figure*} 

SCABS uses the large field-of-view of DECam to image \cena\ out to its approximate virial radius of $\sim300\,{\rm kpc}$, shown in Fig.\,\ref{fig:dither} as the red-dashed ellipse. A five-point dithering strategy per pointing covers the DECam chip gaps, and results in the flower-like mosaic of DECam footprints shown by the blue shading in Fig.\,\ref{fig:dither}. To account for the missing N30 and S30 chips, each pointing overlaps the Northern- and Southern-most CCD rows, and is shifted either E or W by a single chip width, which gives rise to the slightly denser zig-zag regions shown by the darker N--S blue bands on Fig.\,\ref{fig:dither}. The first night of the 2014A-0610 program focussed on the $i'$, $z'$, and $u'$ filters, split into $5\times20=100\,{\rm s}$, $5\times40=200\,{\rm s}$, and $5\times240=1\,200\,{\rm s}$ exposures, respectively. The second night was used to finish the $u'$-band exposures (Tiles\,13--23; see Fig.\,\ref{fig:dither}) and to conduct the short $5\times12=60\,{\rm s}$ and $5\times20=100\,{\rm s}$ $r'$- and $g'$-band SCABS imaging. These exposure times were selected in order to reach targeted apparent point-source magnitudes of $m_{u'}\simeq24.5\,{\rm mag}$, $m_{g'}\simeq23.7\,{\rm mag}$, $m_{r'}\simeq23.1\,{\rm mag}$, $m_{i'}\simeq23.0\,{\rm mag}$, and $m_{z'}\simeq22.8\,{\rm mag}$, corresponding to $3\sigma$ past the GCLF turnover magnitude, assuming a dispersion of $\sigma_{g'}=1.07$ \citep{vil10}. Thus, at these depths up to 99\% of GCs should be detectable in all filters; however, a detailed assessment of the SCABS image quality and limiting magnitudes is deferred to \S\,\ref{sec:comptest}.

Due to intermittent thin clouds combined with a telescope malfunction, nearly the whole of the latter half of the second night of the 2014A-0610 program was lost. For this reason, the $g'$- and $r'$-band observations, as well as a single dither position of the $u'$-band for Tiles 13--23 were not conducted until the 2014B-0609 run. These data are still undergoing reduction before combination with the previous data can be made and thus are not considered further here. Table~\ref{tbl:obs} gives a summary of the data collected, reduced, and analyzed in this work. The observing conditions varied over the course of the two nights with light cirrus giving way to clear and stable conditions during the first night, with a similar trend during the first half of the second followed by deteriorating conditions in the latter half. Seeing as reported by the CTIO {\sc dimm} averaged $0.81\arcsec$ with a dispersion of $0.14\arcsec$ during the first night, while the second night had poorer conditions with mean {\sc dimm} seeing of $1.2\arcsec$. The first half of the second night was also less stable, with a dispersion of $0.34\arcsec$, largely due to a short-term spike in the seeing that particularly affected the photometric calibration of the $r'$-band images (see \S\,\ref{sec:photcalib}).

\begin{table*}
	\centering
	\caption{SCABS observational log. Cols.\,(1) and (2) list the Tile ID (see Fig.~\ref{fig:dither}) and date of observation, followed by the right ascension and declination of the central dither pointings in Cols.\,(3) and (4), respectively. The remaining five columns list total exposure times for, in order, the $u'$, $g'$, $r'$, $i'$, and $z'$ filters.}
	\label{tbl:obs}
	\begin{tabular}{ l c c c c c c c c }
		\hline
		Tile 	& 	Date	&	$\alpha$		&	$\delta$					&	$u'$	&	$g'$	&	$r'$	&	$i'$	&	$z'$	\\
			&		&	($hh:mm:ss$)	&	($^{\circ}$:\arcmin:\arcsec)	&	(s)	&	(s)	&	(s)	&	(s)	&	(s)	\\
		\hline
%		``Central Tile''	\\
%		\hline
		1	&	2014 Apr. 4	&	13:25:27.62	&	-43:01:08.81	&	2\,400	&	...		&	...	&	300		&	400	\\
			&	2014 Apr. 5	&				&				&	...		&	100		&	60	&	...		&	...	\\
		\hline
%		``Outer Ring''	\\
%		\hline
		2 	&	2014 Apr. 4	&	13:32:40.12	&	-41:52:05.25	&	1\,200	&	...		&	...	&	100		&	200	\\
			&	2014 Apr. 5	&				&				&	...		&	100		&	60	&	...		&	...	\\
		3 	&	2014 Apr. 4	&	13:23:46.73	&	-41:12:37.50	&	1\,200	&	...		&	...	&	100		&	200	\\
			&	2014 Apr. 5	&				&				&	...		&	100		&	60	&	...		&	...	\\
		4 	&	2014 Apr. 4	&	13:16:30.83	&	-42:21:41.06	&	1\,200	&	...		&	...	&	100		&	200	\\
			&	2014 Apr. 5	&				&				&	...		&	100		&	60	&	...		&	...	\\
		5 	&	2014 Apr. 4	&	13:18:13.56	&	-44:10:12.36	&	1\,200	&	...		&	...	&	100		&	200	\\
			&	2014 Apr. 5	&				&				&	...		&	100		&	60	&	...		&	...	\\
		6 	&	2014 Apr. 4	&	13:27:12.74	&	-44:49:40.11	&	1\,200	&	...		&	...	&	100		&	200	\\
			&	2014 Apr. 5	&				&				&	...		&	100		&	60	&	...		&	...	\\
		7 	&	2014 Apr. 4	&	13:34:21.17	&	-43:40:36.55	&	1\,200	&	...		&	...	&	100		&	200	\\
			&	2014 Apr. 5	&				&				&	...		&	100		&	60	&	...		&	...	\\		
		\hline
	\end{tabular}
\end{table*}
%
%%%%%%%%%%%%%%%%%%%%%%%%%%%%%%%
%%%%%%%%%%%%%%%%%%%%%%%%%%%%%%%
%%%%%%%%%%%%%%%%%%%%%%%%%%%%%%%
%
\section{Data Reduction}
\label{sec:reduction}
\begin{figure*}
\centering
\includegraphics[width=0.75\linewidth]{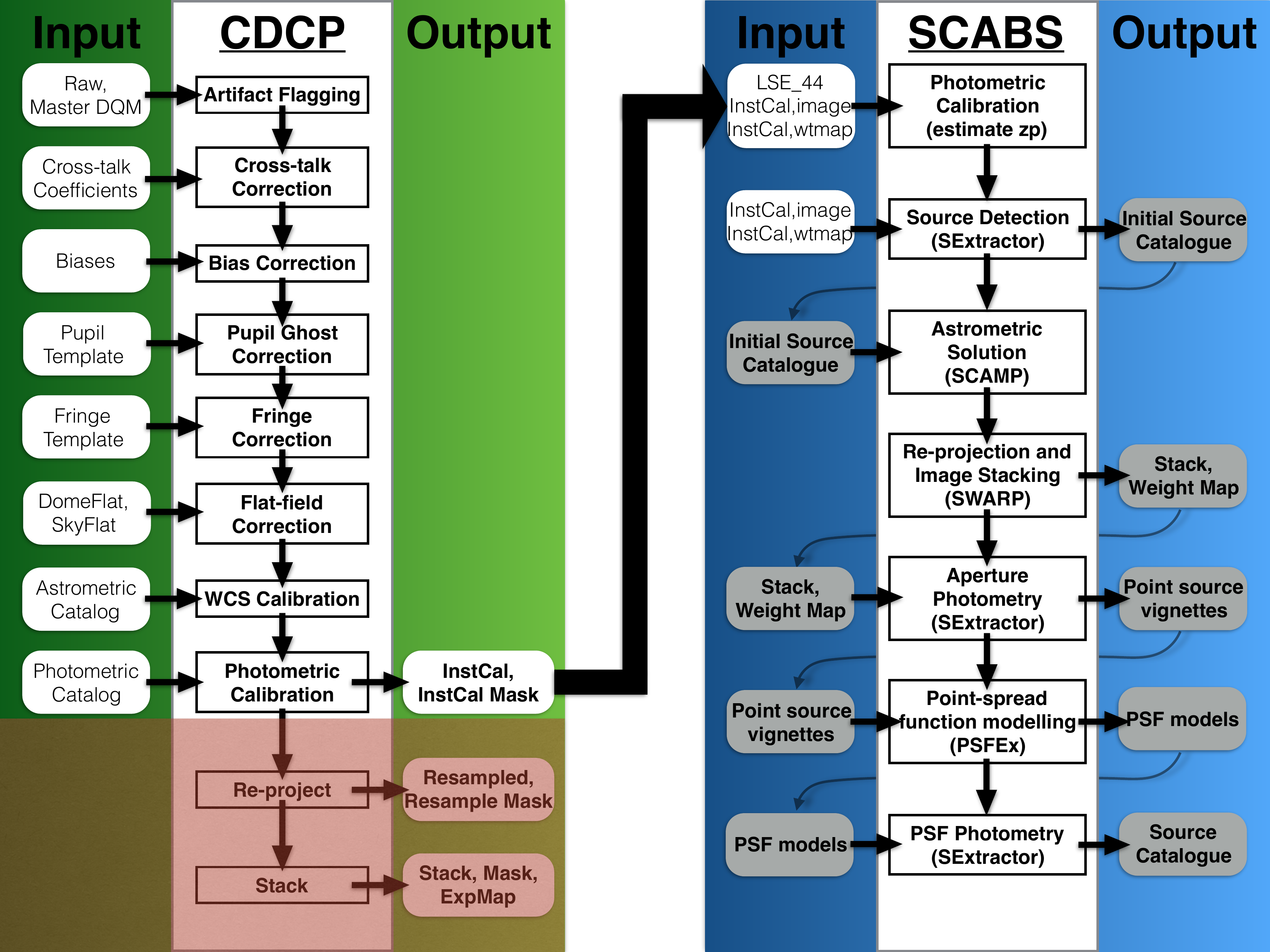}
\caption{The SCABS data reduction cascade. The preliminary CDCP reduction steps are shown on the left (adapted from the DECam User's Guide v.2.0.5) where the {\it InstCal} products are taken as a starting point for our custom reduction steps depicted by the right-hand flowchart, and described in more detail in the text.}
\label{fig:redflow}
\end{figure*}

This section describes the multi-stage processing applied to the raw data taken at the Blanco telescope, which is schematically represented in Fig.\,\ref{fig:redflow}. Briefly, the process includes initial pipeline-based calibrations, from which we take low-level calibrated products and further process them using a custom reduction cascade that derives the final astrometric and photometric solutions, image stacking, and the production of final photometric source catalogues.

%%%%%%%%%%%%%%%%%%%%%%%%%%%%%%%
%%%%%%%%%%%%%%%%%%%%%%%%%%%%%%%
%%%%%%%%%%%%%%%%%%%%%%%%%%%%%%%

\subsection{Pipeline Pre-Processing}
Preliminary reduction steps for all DECam images are performed by the CTIO DECam Community Pipeline \citep[CDCP;][v.3.1.1]{val14} and are shown in more detail by the left-hand flowchart in Fig.\,\ref{fig:redflow} (adapted from the DECam User's Guide v.2.0.5). While in principle the CDCP can provide fully calibrated, sky subtracted, and stacked images, the photometric calibration can be inaccurate by an unknown amount, possibly reaching as high as 0.5\,mag, and thus are not appropriate for our science goals (see the {\it NOAO Data Handbook\footnote{\url{http://ast.noao.edu/sites/default/files/NOAO_DHB_v2.2.pdf}}} for details). For this reason we start with {\it InstCal} products from the CDCP, and perform further reduction/calibration steps using custom routines described below. The {\it InstCal} images have basic pre-processing steps applied to them, correcting for electronic bias, crosstalk between DECam chip amplifiers, and fringing effects. Additionally, the {\it InstCal} frames have been flat-fielded before rudimentary astrometric and photometric calibrations are applied. The end result are frames that are clean of all major cosmetic defects, along with the data quality and background weight maps which are used for further data reduction steps and subsequent analysis as described in the following.

%%%%%%%%%%%%%%%%%%%%%%%%%%%%%%%
%%%%%%%%%%%%%%%%%%%%%%%%%%%%%%%
%%%%%%%%%%%%%%%%%%%%%%%%%%%%%%%

\subsection{Main Image Processing}
Main image processing was carried out by a custom {\sc IDL}- and {\sc Python}-based data reduction pipeline\footnote{\url{https://github.com/rpmunoz/DECam/tree/master/data_reduction}}, which itself calls standard image processing packages from the {\sc Astromatic} software \citep{ber96,ber02,ber06,ber11}. The reduction cascade, as shown by the right-hand schematic in Fig.\,\ref{fig:redflow}, consists of initial source detections made on the individual frames provided by the CDCP, which are then used in the subsequent astrometric and photometric calibrations. Using the astrometric solutions, frames are aligned and combined for each tile--filter combination to produce stacked images suitable for point-source photometry. The photometric measurements are made by performing a second round of source detections on the image stacks, from which bright, unsaturated point sources are identified for point spread function (PSF) modelling. Final photometric catalogues are generated by integrating over the resulting PSF models. These data-reduction cascade steps are listed in detail in the following sub-sections.

%%%%%%%%%%%%%%%%%%%%%%%%%%%%%%%
%%%%%%%%%%%%%%%%%%%%%%%%%%%%%%%
%%%%%%%%%%%%%%%%%%%%%%%%%%%%%%%

\subsubsection{Initial Source Detection}
The first step in constructing the final image stacks is a ``first-pass'' source detection. These sources are used for deriving the astrometric and photometric calibrations described below, and are detected in the individual CDCP {\it InstCal} frames on a chip-by-chip basis using {\sc SExtractor} ({\sc SE}; v.\,2.19.5). For the purpose of calibration, relatively bright, well defined sources are preferred, and so we only took sources detected at a relatively conservative 1.8$\sigma$ above the background ({\sc detect\_thresh=1.8}). To maximize the accuracy of the astrometric solution described below, all CDCP-processed frames are considered at this stage, resulting in 455 individual source catalogues.

%%%%%%%%%%%%%%%%%%%%%%%%%%%%%%%
%%%%%%%%%%%%%%%%%%%%%%%%%%%%%%%
%%%%%%%%%%%%%%%%%%%%%%%%%%%%%%%

\subsubsection{Astrometric Calibration}
The software package {\sc Scamp} (v.\,2.0.4) was used to derive the relative astrometric solution across the SCABS field. {\sc Scamp} reads in the output source catalogues provided by {\sc SE} in the previous step, and matches them to sources from the 2MASS astrometric reference catalogue \citep{skr06}. For this procedure, we set a reference star search radius of $1.2\arcsec$ ({\sc position\_maxerr=1.2}) and only use matches with signal-to-noise ratios (S/N) between 40 and 80 ({\sc sn\_threshholds}=40,80), which typically results in several hundred reference star matches per CCD in a given filter.% This is based on the ``survey\_tX\_dN\_filter\_short\_stars.ldac'' catalogues output by {\sc scamp}}.

\begin{figure}
\centering
\includegraphics[width=\columnwidth,viewport = 164 80 740 558, clip]{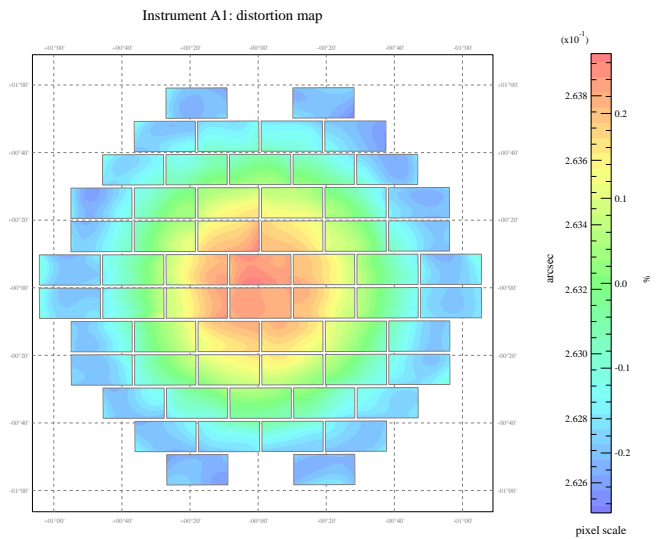}
\caption{The DECam pixel scale distortion map from the astrometric calibration. The field of view of DECam is illustrated with pixel-scale distortion shown by the colour map in units of $''$/pixel, and as an absolute percentage of the mean of 0.263$''$/pixel. Red represents positive pixel-scale distortions, while blue shows distortions smaller than the mean. As a whole, the distortions are very symmetric in nature, with a peak-to-valley difference near the 0.5 percent level.}
\label{fig:distort}
\end{figure}

We apply {\sc lanczos2} resampling to the images, and use a fourth degree polynomial to calculate the pixel distortion across the DECam field-of-view ({\sc distort\_degrees=4}). Fig.\,\ref{fig:distort} shows the pixel scale distortion map resulting from the use of all 455 source catalogues in deriving the astrometric solution. The pixel scale smoothly changes in a radially symmetric way, with distortions of, at most, $\sim0.0015\arcsec$ ($\sim0.5$ percent) between the centre and the edge of the DECam field-of-view. The resulting 1D differences between source coordinates and astrometric reference stars are found to be Gaussian in distribution, centred on $0''$, and with typical $\sigma\lesssim0.2\arcsec$, indicating sub-arcsecond accuracy of our astrometric calibration.

%%%%%%%%%%%%%%%%%%%%%%%%%%%%%%%
%%%%%%%%%%%%%%%%%%%%%%%%%%%%%%%
%%%%%%%%%%%%%%%%%%%%%%%%%%%%%%%

\subsubsection{Photometric Solution}
\label{sec:photcalib}
During the nights we observed $u'g'r'i'z'$ standard star field {\sc LSE\_44} centred at $(\alpha,\delta)=(13h\!:\!52m\!:\!49s,-48^\circ\!:\!09'\!:\!09'')$ \citep{smi07} with varying exposure time and airmass combinations to facilitate photometric zero point estimates. For the estimates we use,
\begin{equation}
\label{eq:zp}
zp=m_{\rm std}-m_{\rm inst}+kX
\end{equation}
where $zp$ is the magnitude zero point, $k$ is the airmass term, $X$ is the airmass, $m_{\rm inst}$ is the instrumental magnitude from the standard star frames based on the {\sc SE} elliptical aperture measurements (i.e.\ {\sc mag\_auto}), and $m_{\rm std}$ is the standard star catalogue AB magnitudes measured in the $u'g'r'i'z'$ system. We estimate $k$ via linear regression in the $\left(m_{\rm std}-m_{\rm inst}\right)$ -- $X$ plane, and then use Eq.\,\ref{eq:zp} to extrapolate the zero point magnitude at $X=1.0$. We compare to zero points and airmass terms based on photometric standard star observations from the period 1--19 November 2012\footnote{\url{http://goo.gl/8h0xuW}}. Table\,\ref{tbl:zp} lists the values derived from our standard star photometry and the CTIO values which are the averages from all DECam CCDs. The listed CTIO errors are those from the CTIO tables, added in quadrature to one standard deviation of the measurements from the 62 CCDs. In general we find very good agreement between the two sets of calibration data, with the exception of the $r'$-band, which differs by $\Delta r'\simeq0.45$\,mag. The image quality during the early part of the second night varied slightly between standard star field exposures, which resulted in a larger scatter in the $m_{r'}$ versus $X$ relation and a correspondingly more uncertain calibration. For this reason we adopt our derived zero points and airmass terms for the four $u'$, $g'$, $i'$, and $z'$ frames, and defer to the older CTIO calibration for the $r'$-band, for which we incorporate errors from both calibrations to reflect the larger calibration uncertainty (see \S\,\ref{sec:phot}--\ref{sec:comptest}).

\begin{table}
	\centering
	\caption{Photometric calibration information. Col.\,(1) lists the filters, while photometric zero points derived directly from our standard star field observations are shown in Col.\,(2), and Col.\,(3) lists those provided by CTIO. Similarly, Cols.\,(4) and (5) list the corresponding airmass terms. The errors on $zp_{\rm CTIO}$ are adopted as the standard deviation between individual chip zero points added in quadrature to the $zp$ error listed in the CTIO calibration data.}
	\label{tbl:zp}
	\begin{tabular}{ l c c c c }
		\hline
		Filter 	& 	$zp_{\rm SCABS}$	&	$zp_{\rm CTIO}$	&	$k_{\rm SCABS}$		&	$k_{\rm CTIO}$	\\
				&	(mag)				&	(mag)			&						&					\\
		\hline
		$u'$		&	23.34$\pm0.06$	&	23.62$\pm0.16$	&	$-$0.32$\pm0.05$	&	$-$0.44$\pm0.03$	\\
		$g'$		&	25.34$\pm0.02$	&	25.42$\pm0.08$	&	$-$0.14$\pm0.02$	&	$-$0.20$\pm0.02$	\\
		$r'$		&	25.02$\pm0.04$	&	25.47$\pm0.06$	&	$-$0.07$\pm0.03$	&	$-$0.10$\pm0.02$	\\
		$i'$		&	25.33$\pm0.01$	&	25.34$\pm0.06$	&	$-$0.02$\pm0.01$	&	$-$0.06$\pm0.01$	\\
		$z'$		&	25.02$\pm0.01$	&	25.06$\pm0.05$	&	$-$0.05$\pm0.01$	&	$-$0.07$\pm0.02$	\\
		\hline
	\end{tabular}
\end{table}

%%%%%%%%%%%%%%%%%%%%%%%%%%%%%%%
%%%%%%%%%%%%%%%%%%%%%%%%%%%%%%%
%%%%%%%%%%%%%%%%%%%%%%%%%%%%%%%

\subsubsection{Image Alignment and Stacking}
\begin{figure}
\centering
\includegraphics[width=\columnwidth]{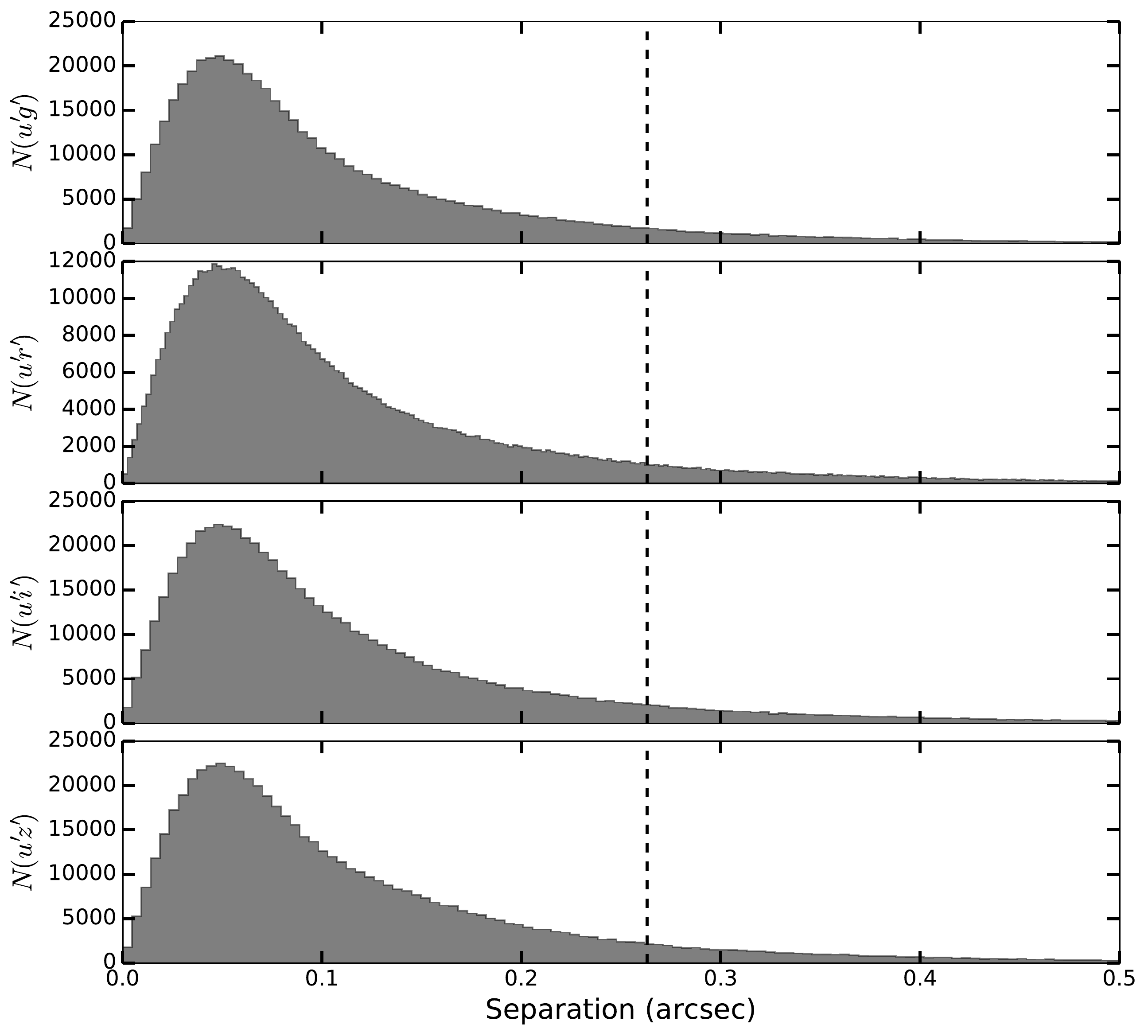}
\caption{Astrometric alignment accuracy. The distributions in source coordinate offsets for recovered sources in Tile\,1 are shown with respect to their $u'$-band astrometry and demonstrate that our image alignments are accurate to within $\sim0.05\arcsec$. Vertical black dashed lines indicate the DECam pixel scale.}
\label{fig:separation}
\end{figure} 

The astrometric solution derived on all the individual frames is used to construct the final image stacks using the {\sc Swarp} software (v.\,2.38). During the initial source detection algorithm, we use the $i'$-band images as our reference frames, and align the others to these images in pixel-space. We find our image alignment to be excellent, as matching recovered sources in each filter in Tile\,1 to the $u'$-band source catalogue results in Fig.\,\ref{fig:separation}, which shows the distribution in source coordinate offsets. In each panel the offset distributions peak at $\sim0.05\arcsec$, with virtually no sources showing offsets of $\gtrsim0.5\arcsec$.

As noted in \S\,\ref{sec:obs}, we failed to finish a complete dither pattern for the $u'$-band observations, and could not obtain data of sufficient quality in the $g'$ and $r'$ filters for Tiles 13-23 of the SCABS footprint. For this reason they were omitted from the data reduction cascade described above and thus no image stacks are available to be analyzed. With this unfortunate fact in mind, and not wanting to delay the release of our current science-ready source catalogues, we restrict this release to sources in Tiles\,1--7, roughly corresponding to the spatially homogeneous halo within $\sim140\,{\rm kpc}$ of \cena. While Tiles\,8--12, and 24 are in principle ready, we will release those data in a future release along with the results of the analysis of Tiles\,13--24.

%%%%%%%%%%%%%%%%%%%%%%%%%%%%%%%
%%%%%%%%%%%%%%%%%%%%%%%%%%%%%%%
%%%%%%%%%%%%%%%%%%%%%%%%%%%%%%%

\subsection{Photometry}
\label{sec:phot}
\begin{table*}
	\centering
	\caption{Point-spread function modelling summary. Col.\ (1) indicates the tile over which the PSF modelling was conducted, followed by the number of PSF stars selected for the $u'$-band in Col.\,(2), and for the $g'$-, $r'$-, $i'$-, and $z'$-bands in each second column thereafter. Similarly, Col.\,(3) lists the PSF FWHM measured in the $u'$-band for each tile, with every second column following lists the same information for the remaining filters. Finally, the bottom row lists the mean numbers of PSF stars, and the corresponding mean FWHM values.}
	\label{tbl:psfmodel}
	\begin{tabular}{lcccccccccc} 
		\hline
		Tile 	& 	$N_{u', {\rm psf}}$	&	$u'_{\rm fwhm}$	&	$N_{g', {\rm psf}}$	&	$g'_{\rm fwhm}$	&	$N_{r', {\rm psf}}$	&	$r'_{\rm fwhm}$	&	$N_{i', {\rm psf}}$		&	$i'_{\rm fwhm}$	&	$N_{z', {\rm psf}}$	&	$z'_{\rm fwhm}$	\\
			&					&	(\arcsec)			&					&	(\arcsec)			&					&	(\arcsec)		&						&	(\arcsec)		&				&	(\arcsec)		\\
		\hline
		1		&	6\,007	&	1.43		&	8\,129	&	1.68		&	10\,806	&	1.27		&	11\,671	&	1.13	&	13\,783	&	1.28	\\
		2		&	6\,736	&	1.43		&	7\,132	&	1.74		&	11\,223	&	1.25		&	8\,154	&	1.43	&	9\,971	&	1.34	\\
		3		&	6\,997	&	1.41		&	9\,101	&	1.68		&	10\,141	&	1.25		&	11\,109	&	1.35	&	10\,533	&	1.28	\\
		4		&	7\,293	&	1.42		&	8\,134	&	1.66		&	10\,899	&	1.26		&	11\,557	&	1.22	&	10\,742	&	1.22	\\
		5		&	8\,282	&	1.41		&	9\,049	&	1.69		&	11\,201	&	1.25		&	12\,027	&	1.25	&	9\,887	&	1.23	\\
		6		&	8\,613	&	1.40		&	9\,812	&	1.85		&	11\,846	&	1.24		&	10\,284	&	1.27	&	10\,192	&	1.28	\\
		7		&	8\,668	&	1.37		&	11\,638	&	1.74		&	11\,896	&	1.25		&	11\,458	&	1.41	&	11\,625	&	1.38	\\
		\hline
		Mean	&	7514		&	1.41		&	8999		&	1.72		&	11145	&	1.25		&	10894	&	1.29	&	10962	&	1.29	\\
		\hline
	\end{tabular}
\end{table*}

For all of the final stacked frames, photometric measurements on detected sources are performed using a combination of {\sc SE} and {\sc PSFEx} (v.\,3.16.1). Sources are first detected with an initial pass of {\sc SE}, and vignette images are created upon which the PSF is modelled by {\sc PSFEx}. Before the PSF modelling, point-like sources are identified on the images by identifying the stellar locus in {\sc mag\_auto}--{\sc flux\_radius} space, and bright but non-saturated sources are selected that bracket the mean {\sc flux\_radius} of the locus. We model the PSF on the corresponding vignettes with {\sc PSFEx} using a $47\times47\,{\rm pix}$ kernel ({\sc psf\_size=47,47}) with variations followed to third order ({\sc psfvar\_degrees=3}). A summary of the modelling is shown in Table\,\ref{tbl:psfmodel} that lists, for each tile--filter combination, the number of PSF models constructed, as well as the modelled PSF FWHM in arcsec, which varies between $1.13\arcsec$ and $1.85\arcsec$, with the best quality observations corresponding to the $r'$-band, which shows a mean FWHM of $1.25\arcsec$.

The resulting PSF models are finally used in another {\sc SE} run, which using them as input, measures the final, PSF-corrected magnitudes for the sources. During this final source extraction, the detection criterion was such that flux from at least six pixels adjacent to a source ({\sc detect\_minarea=6}) fell at least 1.5$\sigma$ above the background levels to be analyzed ({\sc detect\_thresh=1.5}, {\sc analysis\_thresh=1.5}). Table\,\ref{tbl:sources} summarizes the final source catalogues produced by {\sc SE} for each stacked frame, with totals throughout the area covered by Tiles\,1--7 listed below. We adopt statistical photometric errors as those reported by {\sc PSFEx}, and add the uncertainties from our adopted zero point and airmass term calibrations (see \S\ref{sec:photcalib}, and Table \ref{tbl:zp}) in quadrature to the 1$\sigma$ bootstrapped uncertainties based on our artificial star experiments (\S\,\ref{sec:comptest}) to our systematic error budget.

\begin{table}
	\centering
	\caption{Source detection summary. The seven tiles are listed in Col.\,(1), while Cols.\,2--6 show the total number of sources with PSF-corrected photometric measurements in each of the $u'$-, $g'$-, $r'$-, $i'$-, and $z'$-band stacked images.}
	\label{tbl:sources}
	\begin{tabular}{lrrrrrr}
		\hline
		Tile 	& 	$N_{u'}$	&	$N_{g'}$		&	$N_{r'}$	&	$N_{i'}$	&	$N_{z'}$	\\%&	$N_{\rm all}$	&	$N_{\rm point}$	&	$N_{\rm GC}$	\\
		\hline
		1	&	91\,287	&	102\,519	&	152\,247		&	286\,018		&	234\,128		\\%&	77\,336	&	68\,434	&	761		\\
		2	&	77\,274	&	95\,236	&	147\,212		&	176\,099		&	175\,538		\\%&	67\,349	&	58\,167	&	362		\\
		3	&	68\,662	&	87\,405	&	130\,956		&	169\,513		&	169\,371		\\%&	60\,627	&	51\,336	&	300		\\
		4	&	73\,094	&	92\,028	&	139\,833		&	191\,969		&	195\,177		\\%&	64\,316	&	51\,876	&	256		\\
		5	&	88\,133	&	108\,055	&	162\,966		&	218\,117		&	201\,702		\\%&	77\,528	&	66\,780	&	344		\\
		6	&	101\,931	&	112\,941	&	185\,733		&	242\,566		&	234\,523		\\%&	87\,345	&	74\,861	&	327		\\
		7	&	94\,755	&	111\,664	&	170\,352		&	206\,237		&	205\,738		\\%&	82\,348	&	67\,495	&	326		\\
		\hline
		All	&	595\,136	&	709\,848	&	1\,089\,299	&	1\,490\,519	&	1\,416\,177	\\%&	516\,849	&	463\,774	&	2\,676	\\
		\hline
	\end{tabular}\\
\end{table}

\begin{figure}
\centering
\includegraphics[width=\linewidth]{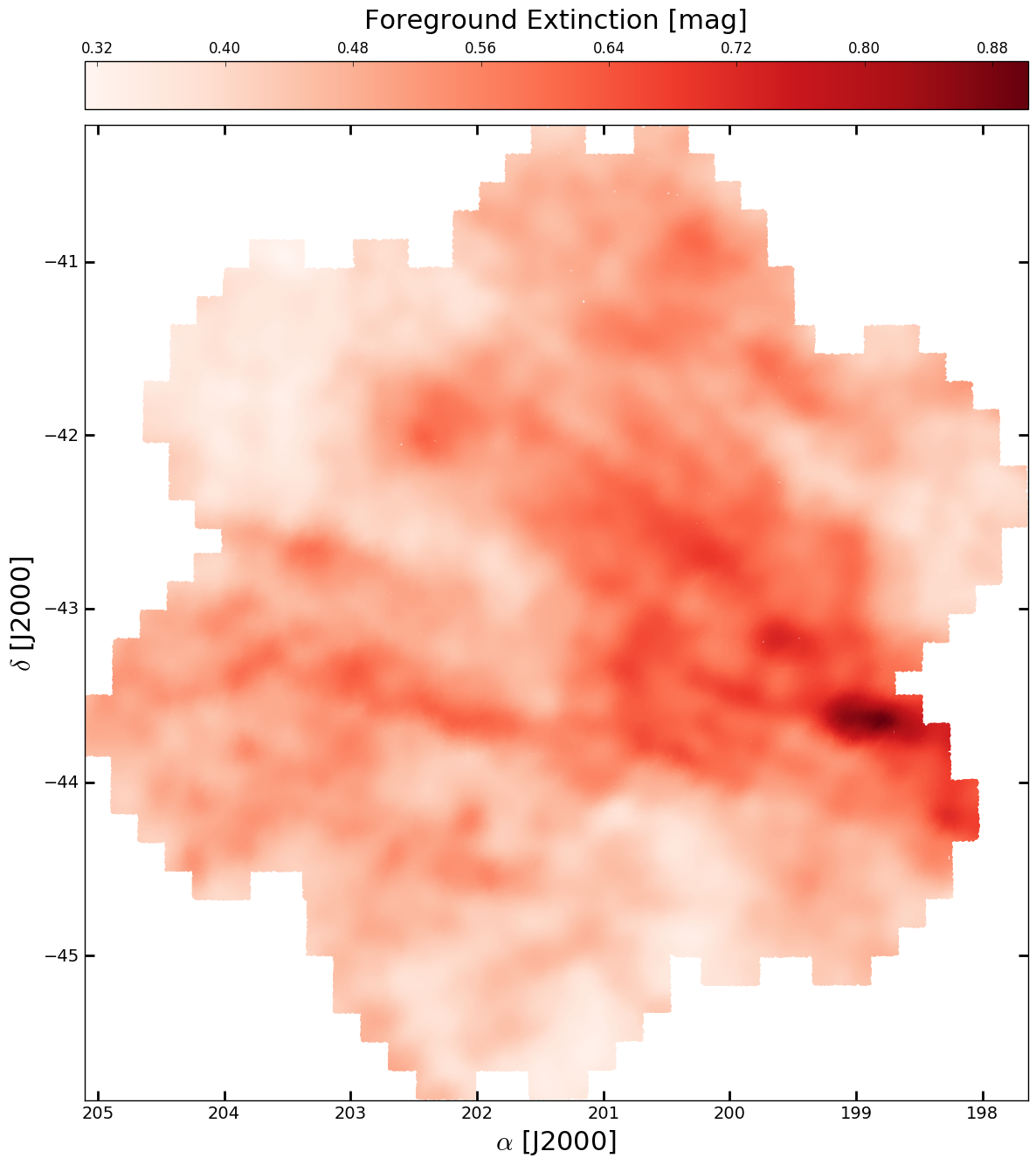}
\caption{The $u'$-band foreground extinction ($A_{u'}$) map for Tiles\,1--7 of SCABS. The amount of extinction across the field of view differs by as much as $0.5$\,mag in the $u'$-band across the field of view, and illustrates the importance of correcting for differential foreground reddening.}
\label{fig:reddening}
\end{figure} 

While no attempt is made to correct for any reddening intrinsic to \cena, the wide field-of-view of DECam necessitates a careful correction for foreground Galactic extinction. To this end, we query the {\it Galactic Extinction and Reddening Calculator}\footnote{\url{http://ned.ipac.caltech.edu/forms/calculator.html}} using the Galactic reddening maps of \cite{sch11} on a source-by-source basis and include these for convenience in the source catalogues described in \S\,\ref{sec:disc} and summarized in Table\,\ref{tbl:source_cats}. Fig.\,\ref{fig:reddening} illustrates the importance of this task, particularly in the $u'$-band, to take into account differential reddening across the SCABS field-of-view. From the heat-map, which indicates the magnitude of reddening towards a given direction in Tiles\,1--7, extinction can be as high as $A_{u'}=0.9\,{\rm mag}$ in the $u'$-band, with peak-to-valley differences across the region of as much as $\Delta A_{u'}=0.59\,{\rm mag}$, with effects diminishing towards the redder filters to a maximum $A_{z'}=0.27\,{\rm mag}$ and $\Delta A_{z'}=0.18\,{\rm mag}$ for the $z'$-band. 

%%%%%%%%%%%%%%%%%%%%%%%%%%%%%%%
%%%%%%%%%%%%%%%%%%%%%%%%%%%%%%%
%%%%%%%%%%%%%%%%%%%%%%%%%%%%%%%

\subsection{Data Quality Assessment}
\label{sec:comptest}
Artificial star experiments were conducted to quantify the point-source depths of the SCABS observations. In these experiments, artificial point-sources are added to images in a range of magnitudes, and the same source detection algorithm used on the science images are applied to the experimental images. The results are used to derive the magnitude limit at which we fail to recover the mock sources.

We use a set of {\sc IDL}- and {\sc Python}-based scripts to add stars by slicing images into $200\times200\,{\rm pix}^2$ regions, and generating up to 10 random positions within each to add the mock stars. To avoid artificial crowding, the positions are such that no real source lies within 16 pixels of the mock star, which results in $\sim60\,000$ artificial star positions per image. Magnitudes are then randomly assigned to the artificial sources to create the final mock star catalogues. To obtain sufficient statistics, the process of assigning random magnitudes is repeated 10 times per image, so that the results are in reality based on $\sim600\,000$ artificial sources per pointing.

\begin{figure}
\centering
\includegraphics[width=\columnwidth]{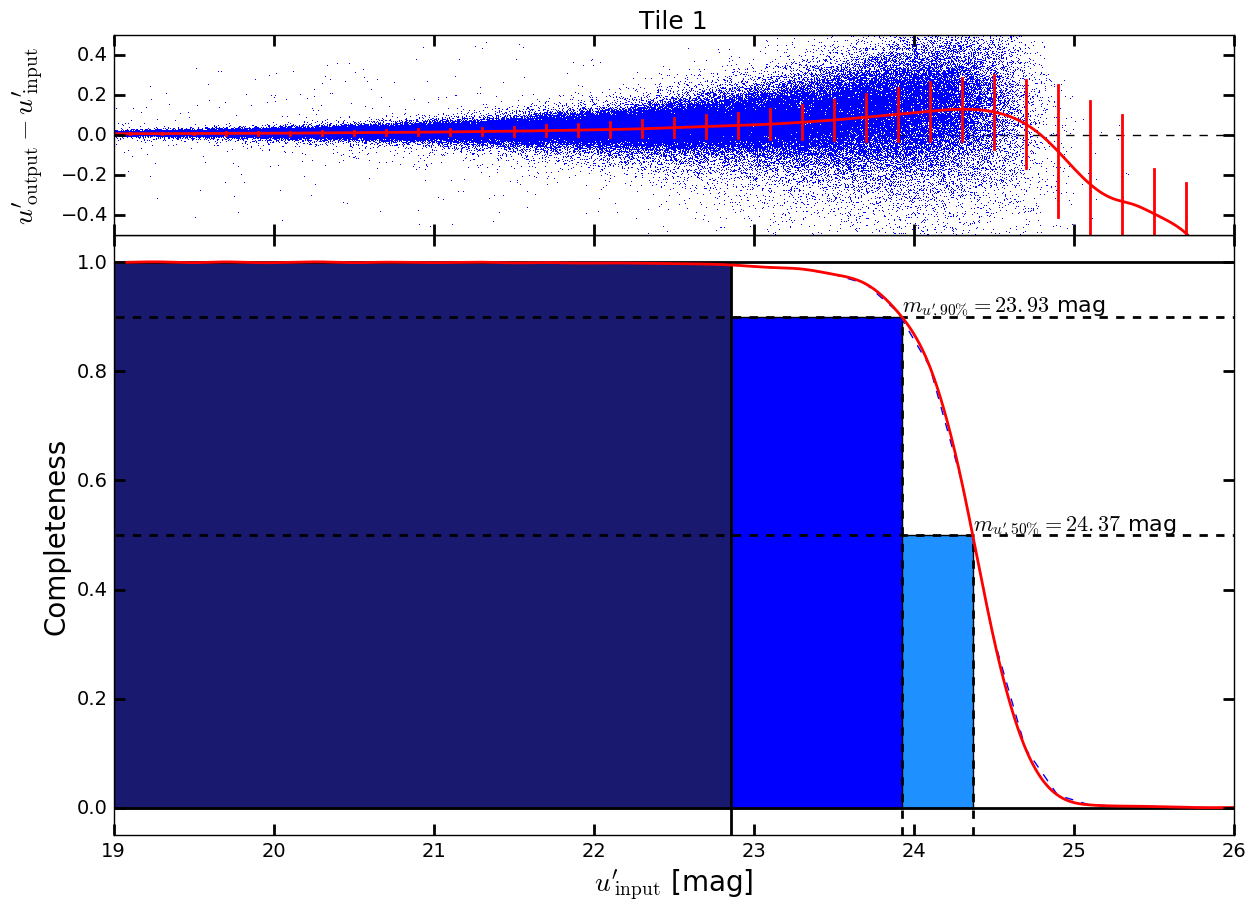}
\includegraphics[width=\columnwidth]{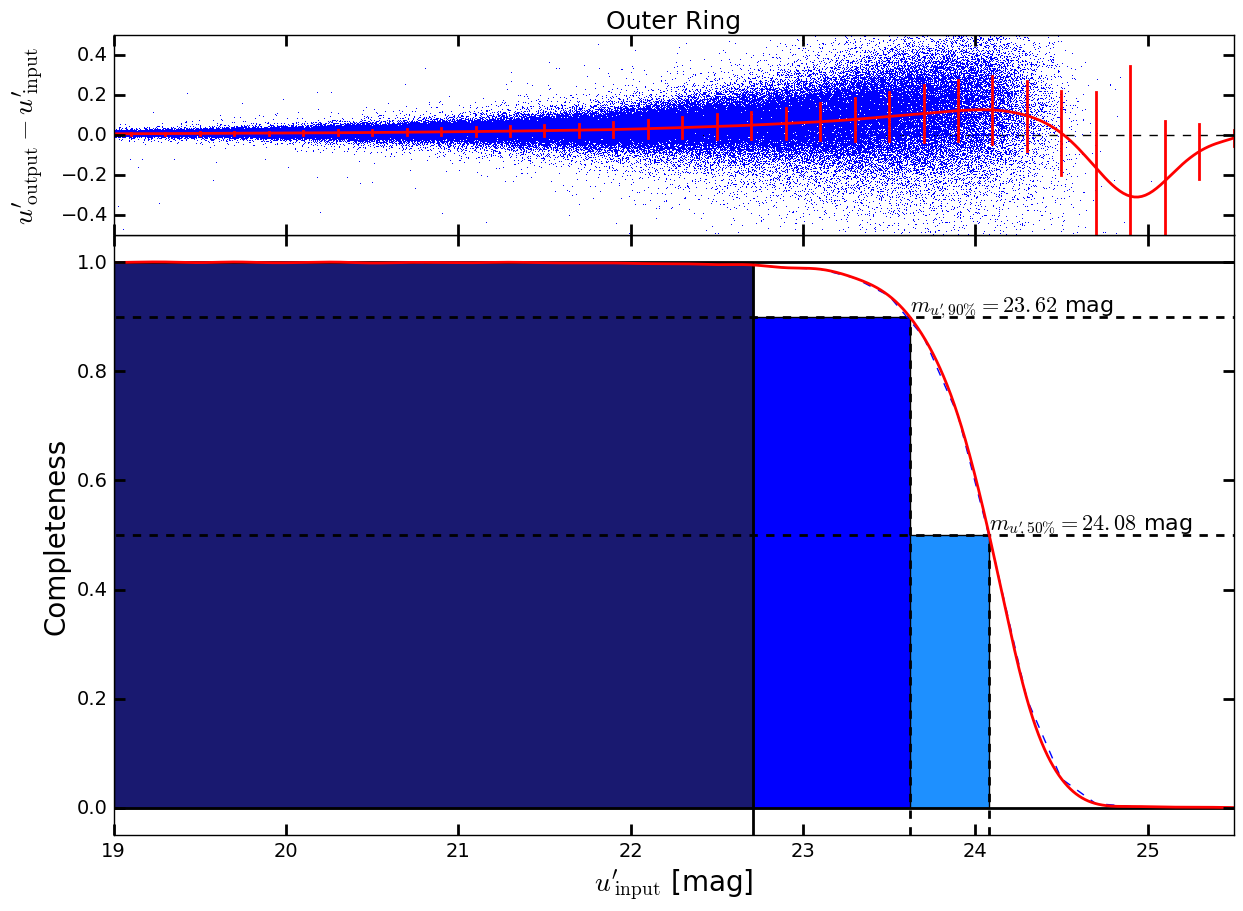}
\caption{Results of our artificial star experiments expressed as $u'$-band point-source completeness estimates. ({\it Upper sub-panels:})\,The difference between the input mock stellar catalogue magnitudes and the output {\sc SE} measurements as a function of input magnitude are shown in blue, while the red curve indicates a non-parametric linear regression to the data, along with 1$\sigma$ bootstrapped errors in 0.2\,mag bins, clipped to within 2.3$\sigma$. The black dashed line indicates perfect agreement between input and output magnitude. ({\it Lower sub-panels:})\,The fractions of sources recovered from the input catalogues in 0.2\,mag bins. The dashed blue line shows the data, while the solid red line shows a spline interpolation used to predict the 50\% and 90\% completeness limits. The blue shading encloses sources with 100\% representation (dark blue), 90\% representation (blue), and 50\% representation (light blue). Adopted 50\% and 90\% completeness limits are shown (see also Tbl.\,\ref{tbl:ch2_complete}. Finally, the top panel shows results corresponding to the central tile (Tile\,1) of SCABS, while the bottom panel shows results representative of the outer ring (Tiles\,2--7).}
\label{fig:comptest_u}
\end{figure} 

\begin{figure}
\centering
\includegraphics[width=\columnwidth]{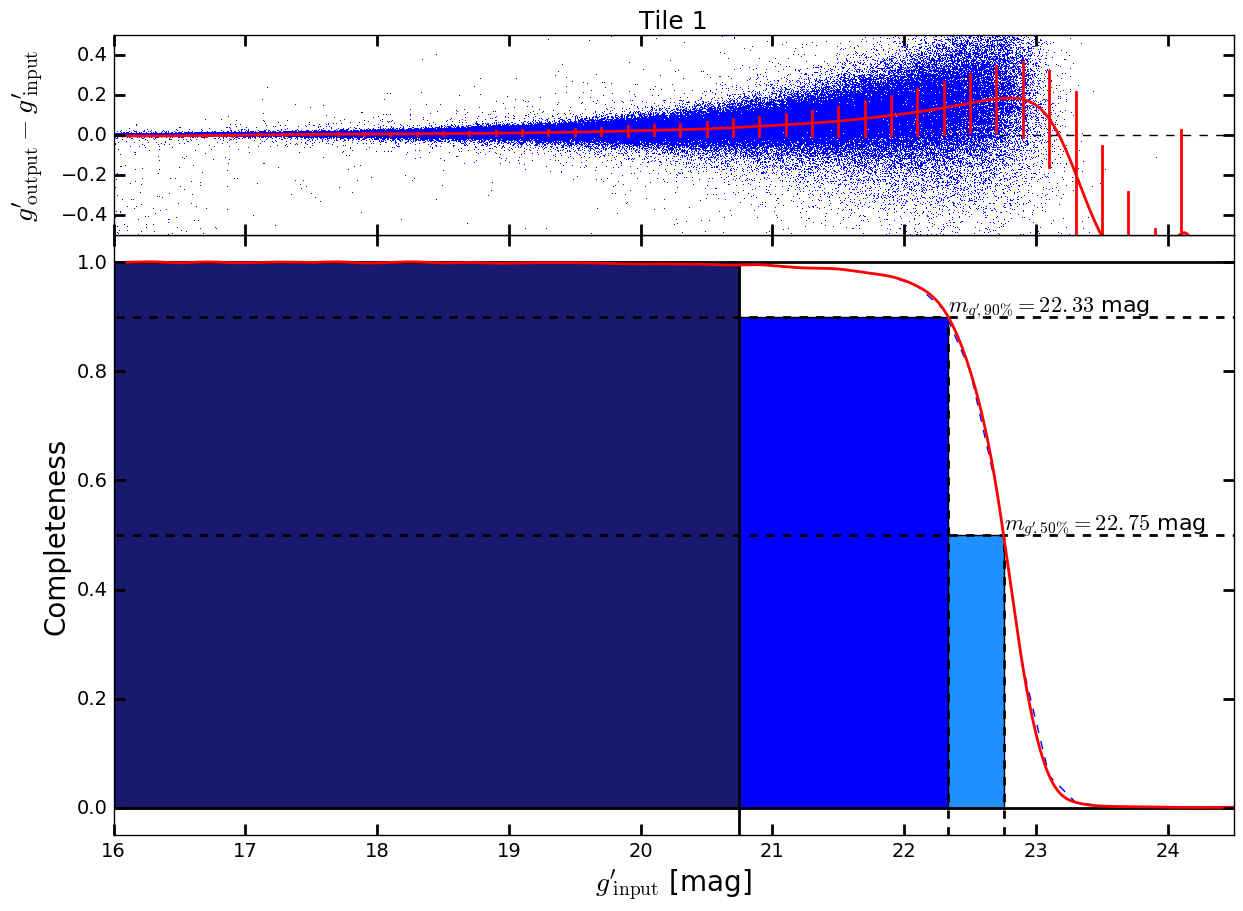}
\includegraphics[width=\columnwidth]{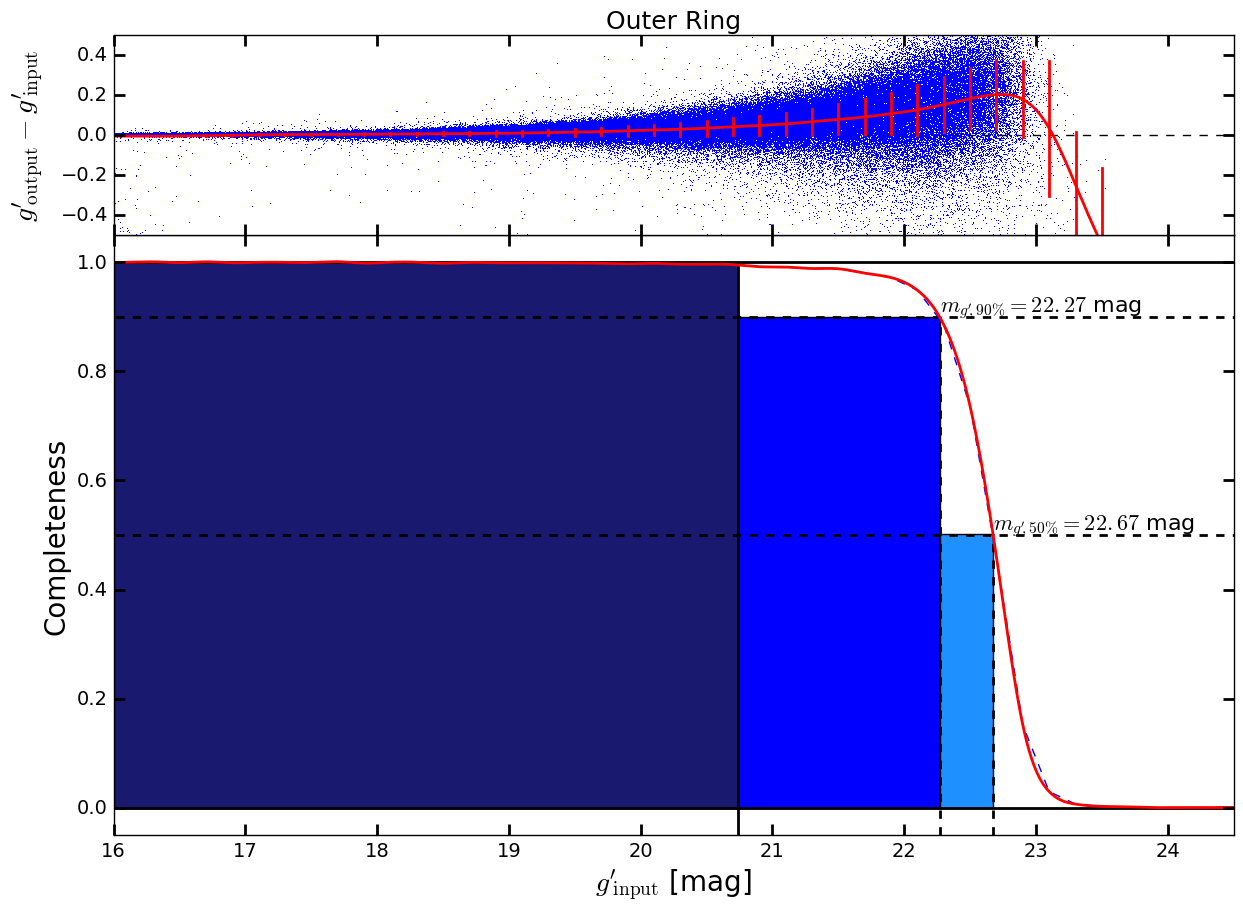}
\caption{Results of the artificial star experiments expressed as $g'$-band point-source completeness estimates. See Fig.\,\ref{fig:comptest_u} for detailed descriptions of the panels.}
\label{fig:comptest_g}
\end{figure} 

\begin{figure}
\centering
\includegraphics[width=\columnwidth]{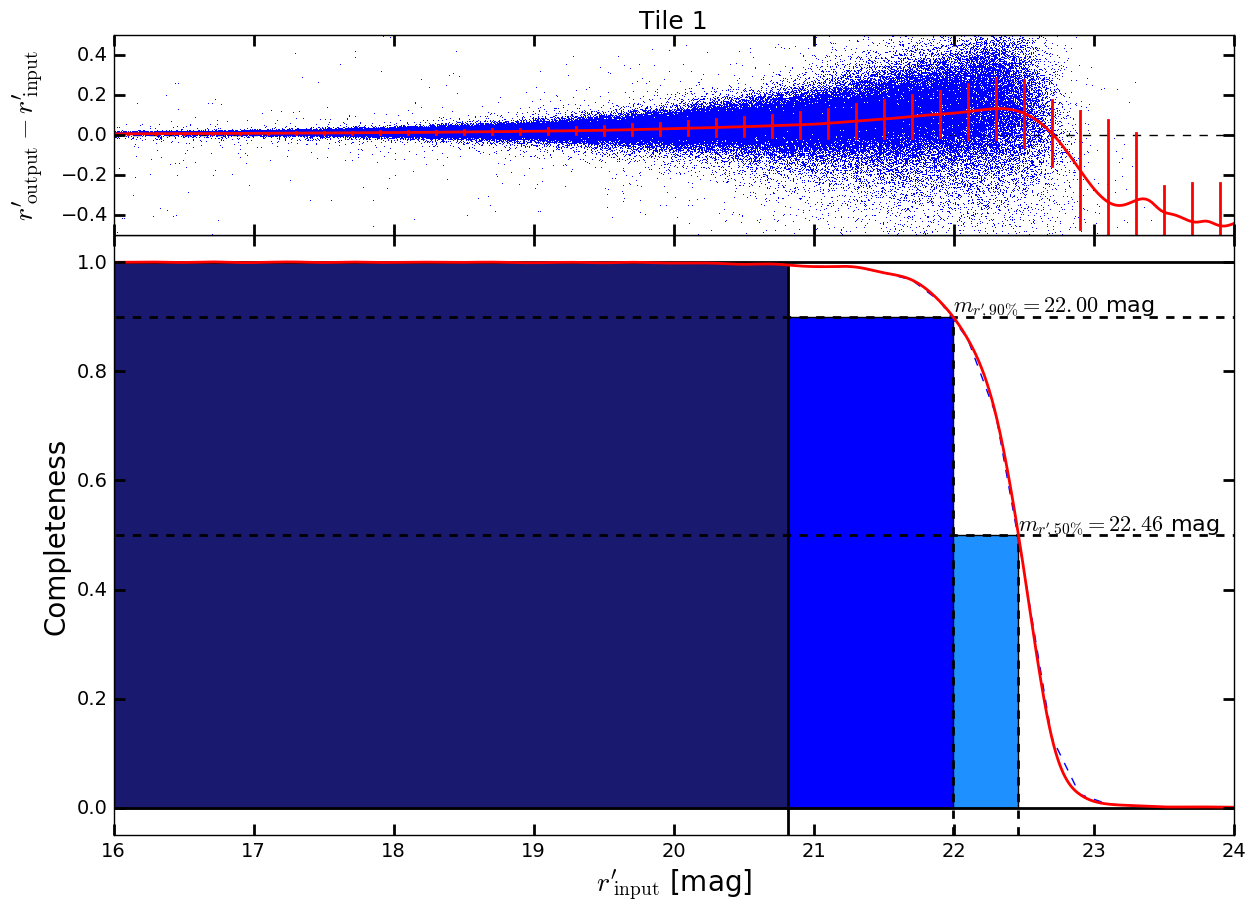}
\includegraphics[width=\columnwidth]{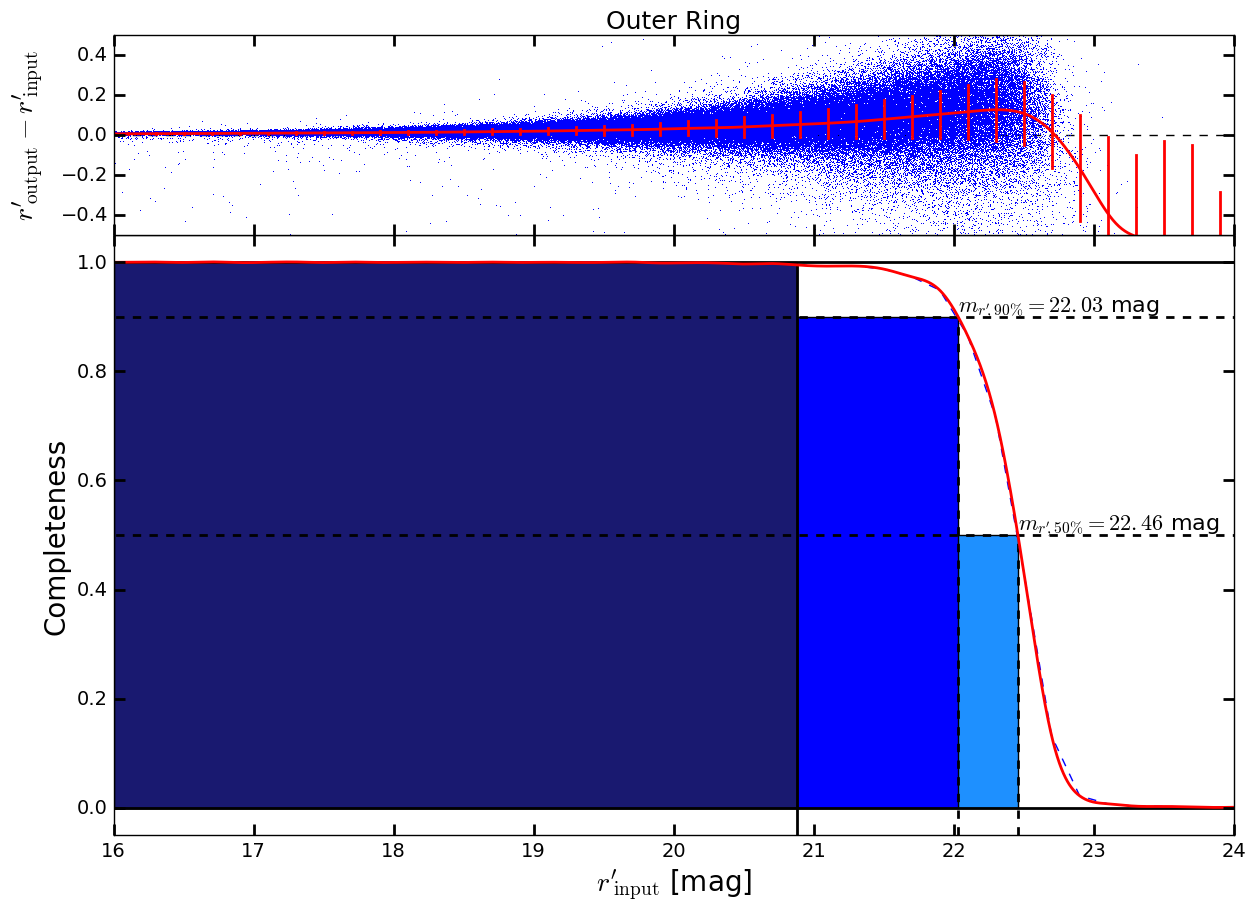}
\caption{Results of the artificial star experiments expressed as $r'$-band point-source completeness estimates. See Fig.\,\ref{fig:comptest_u} for detailed descriptions of the panels.}
\label{fig:comptest_r}
\end{figure} 

\begin{figure}
\centering
\includegraphics[width=\columnwidth]{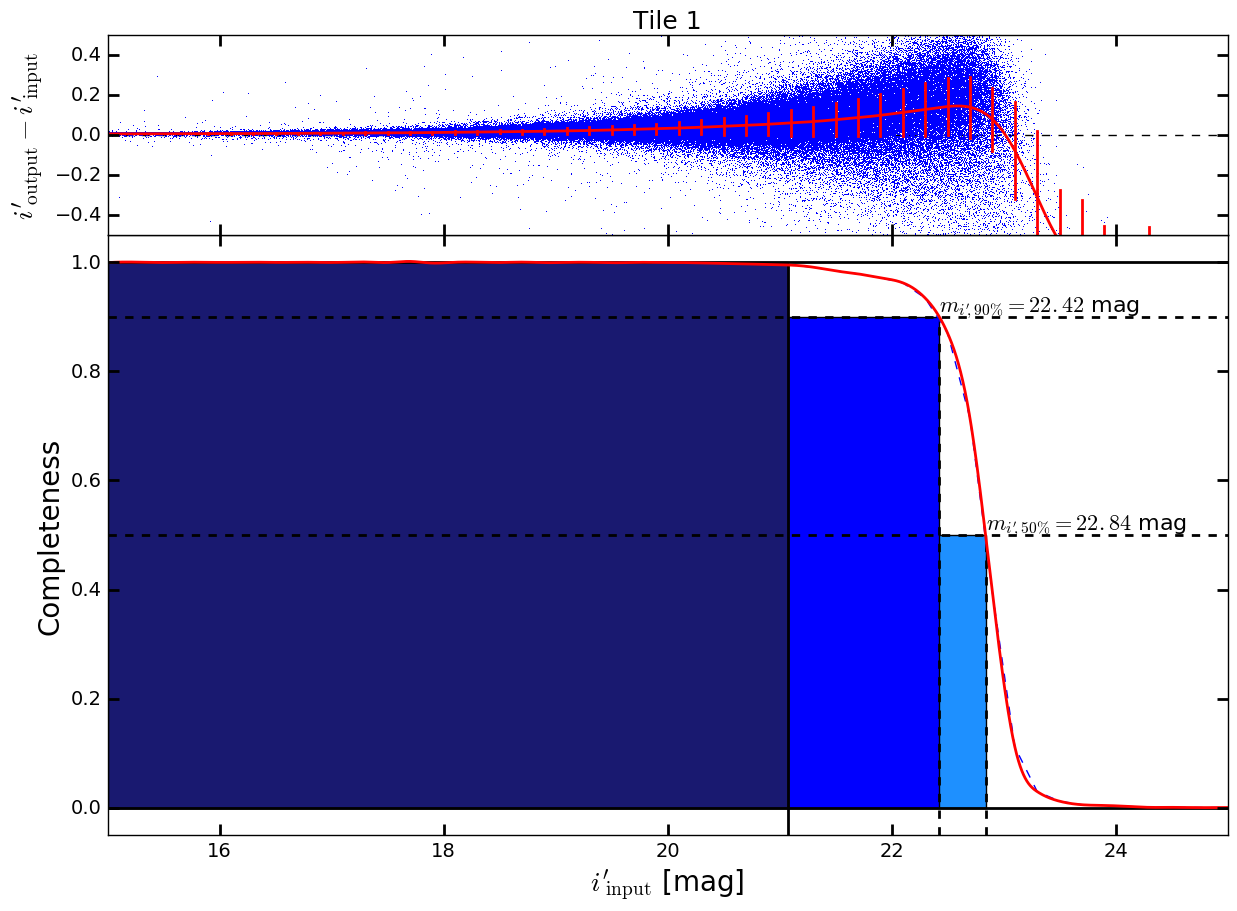}
\includegraphics[width=\columnwidth]{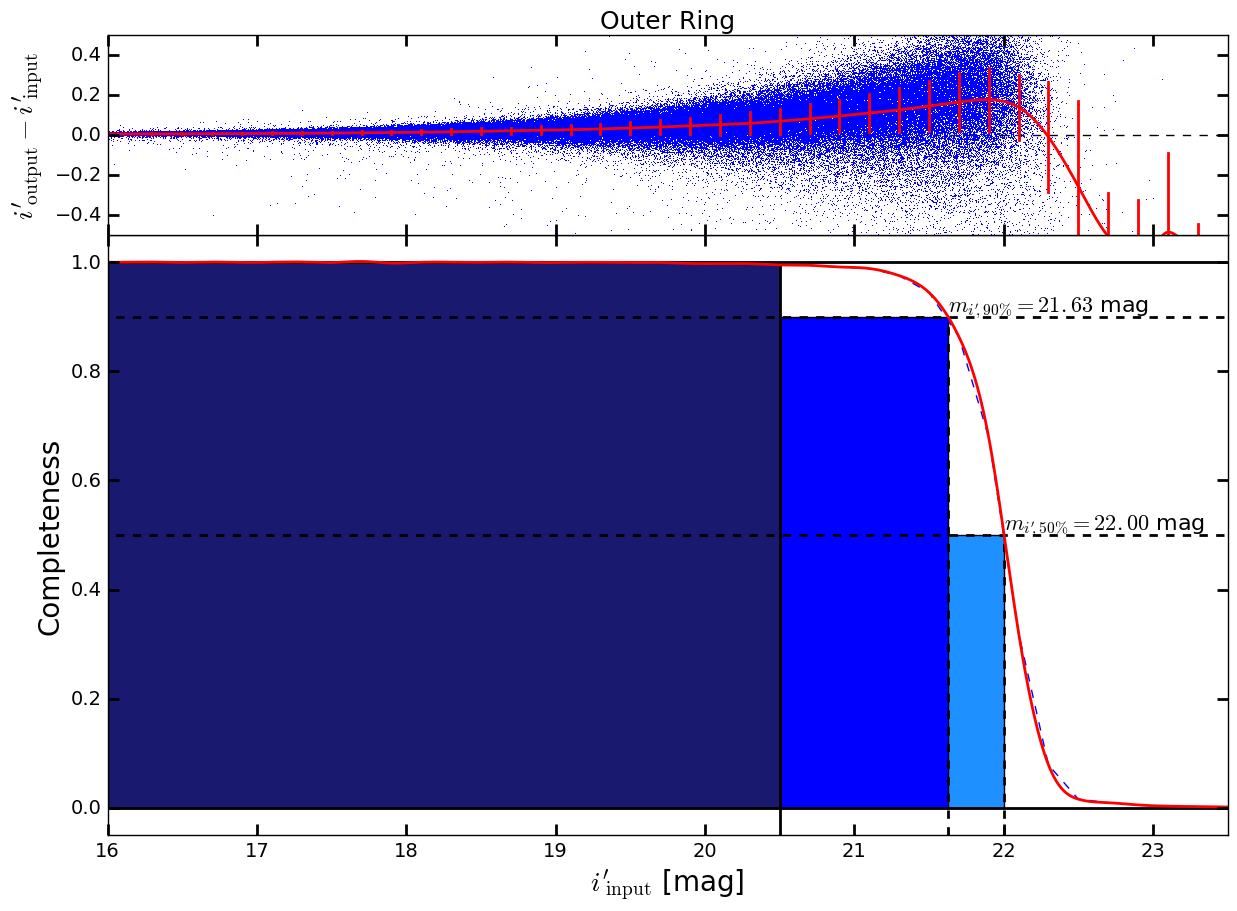}
\caption{Results of the artificial star experiments expressed as $i'$-band point-source completeness estimates. See Fig.\,\ref{fig:comptest_u} for detailed descriptions of the panels.}
\label{fig:comptest_i}
\end{figure} 

\begin{figure}
\centering
\includegraphics[width=\columnwidth]{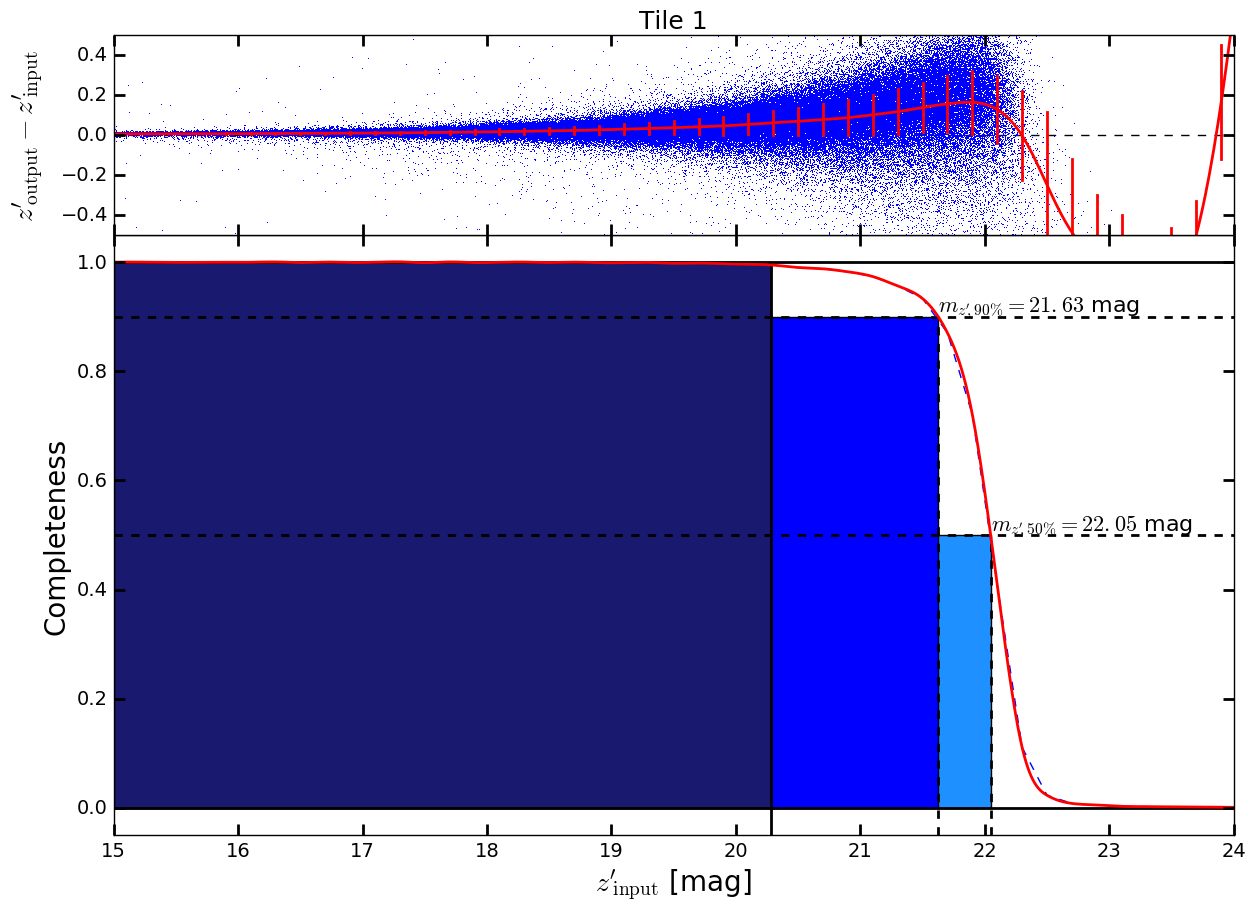}
\includegraphics[width=\columnwidth]{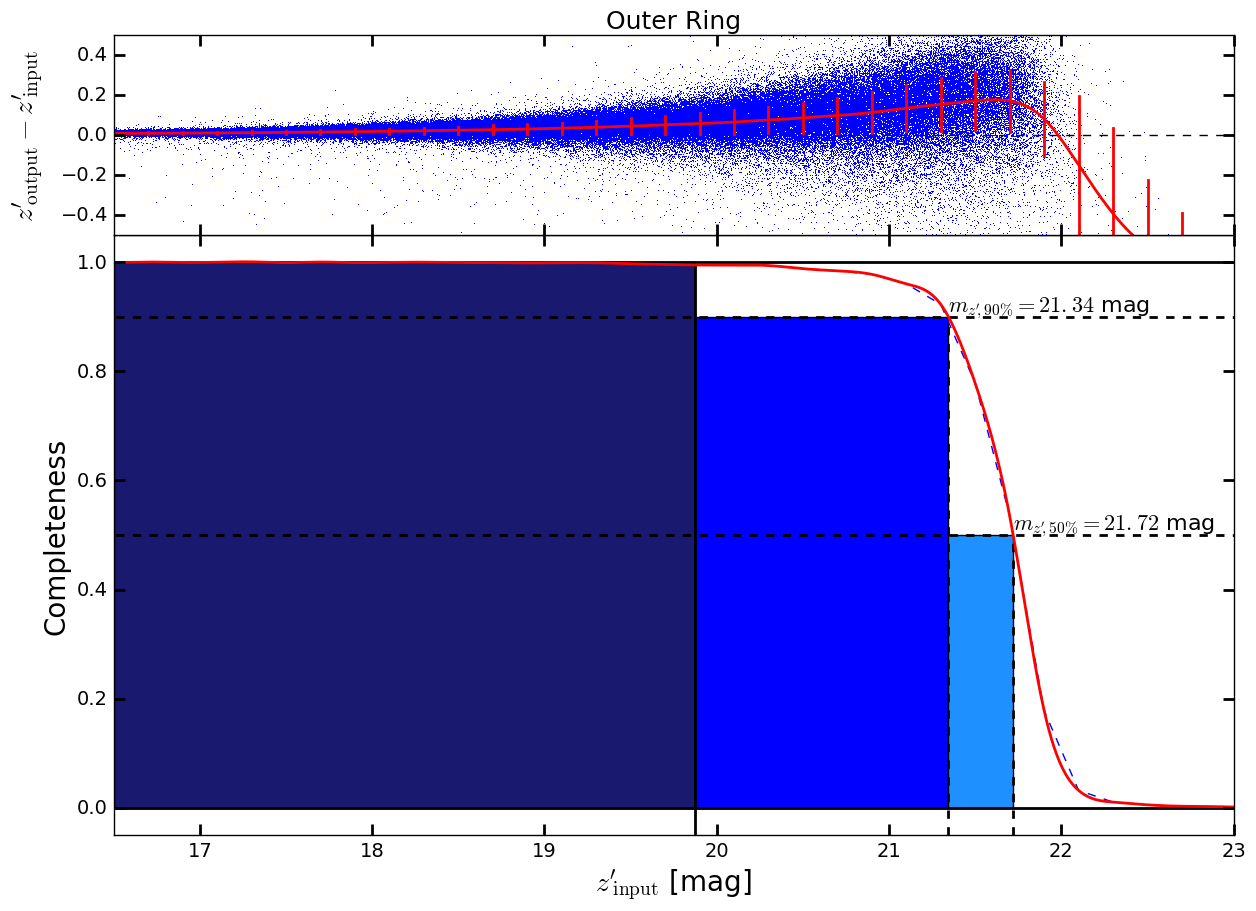}
\caption{Results of the artificial star experiments expressed as $z'$-band point-source completeness estimates. See Fig.\,\ref{fig:comptest_u} for detailed descriptions of the panels.}
\label{fig:comptest_z}
\end{figure} 

\begin{table}
	\centering
	\caption{Point source completeness estimates.  The Table is separated into two subtables indicated by the subtitles. The first five rows list completeness limits for the central tile, where $u'$, $i'$, and $z'$ photometry is deeper than the rest of the survey, while the bottom five rows list values that are assumed for Tiles\,2--7 (i.e.\,"the Outer Ring"). Col.\,(1) indicates the filter under consideration, while Cols.\,(2) and (4) show the 50 and 90 percent completeness limits, respectively. Cols.\,(3) and (5) show one standard deviation from the mean of the 50 and 90 percent completeness magnitudes across the DECam field of view, which is an indication of the spatially varying photometric sensitivity of SCABS.}
	\label{tbl:complete}
	\begin{tabular}{lcccc}
		\hline
		Filter 	& 	$m_{50\%}$	&	$\sigma_{50\%}$	&	$m_{90\%}$	&	$\sigma_{90\%}$	\\
				&	(mag)		&	(mag)			&	(mag)		&	(mag)			\\
		\hline
		Tile\,1\\
		\hline
		$u'$		&	24.37	&	0.25	&	23.93	&	0.30	\\
		$g'$		&	22.75	&	0.20	&	22.33	&	0.20	\\
		$r'$		&	22.46	&	0.17	&	22.00	&	0.26	\\
		$i'$		&	22.84	&	0.22	&	22.42	&	0.31	\\
		$z'$		&	22.05	&	0.19	&	21.62	&	0.29	\\
		\hline
		Tiles\,2--7\\
		\hline
		$u'$		&	24.08	&	0.25	&	23.62	&	0.25	\\
		$g'$		&	22.67	&	0.20	&	22.27	&	0.20	\\
		$r'$		&	22.46	&	0.24	&	22.03	&	0.20	\\
		$i'$		&	22.05	&	0.24	&	21.63	&	0.19	\\
		$z'$		&	21.72	&	0.14	&	21.34	&	0.27	\\
		\hline
	\end{tabular}
\end{table}

We simulate the SCABS observing conditions by querying our PSF model libraries (see \S\,\ref{sec:phot}) and scaling them in luminosity to add appropriate ``stellar'' sources at the corresponding positions to the real image. Another run of {\sc SE} using the same parameters as for the real science frames is carried out on the experimental images, and mock output catalogues are generated. These catalogues are compared to the input mock catalogues to derive the completeness limits of the SCABS data. To account for the longer exposure times in the $u'$, $i'$, and $z'$ filters for Tile\,1, this process is carried out separately for all five filters in Tiles\,1 and 2. Given the constant exposure times, FWHM dispersions not exceeding $\sim8$ percent relative to the mean in any band, and the good agreement between the Tile\,1 and 2 $g'$- and $r'$-band sensitivities (see Table\,\ref{tbl:complete}), the results of the artificial star experiments for Tile\,2 are adopted for Tile\,3--7.

\begin{figure*}
\centering
\includegraphics[width=0.49\linewidth]{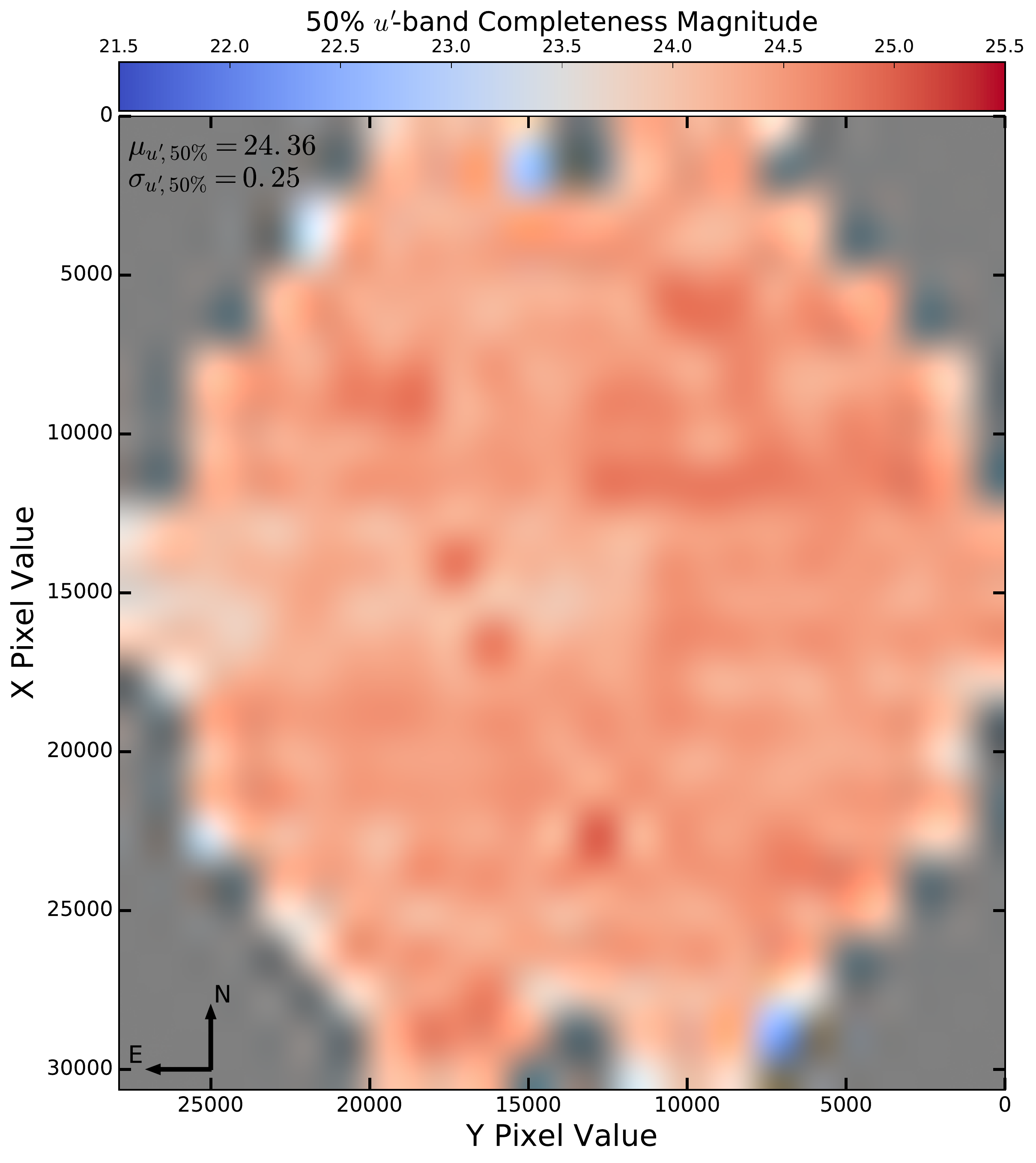}
\includegraphics[width=0.49\linewidth]{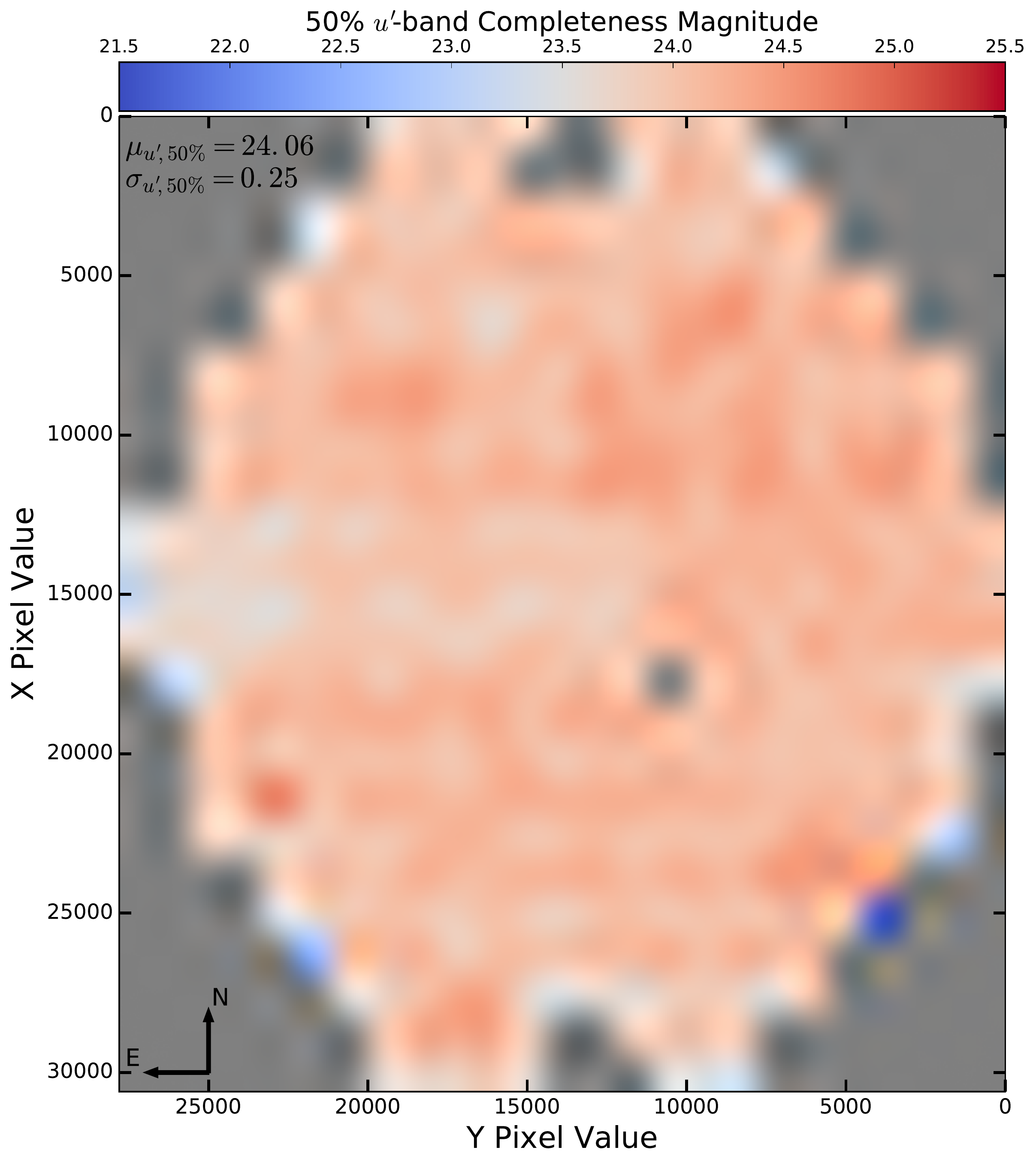}
\includegraphics[width=0.49\linewidth]{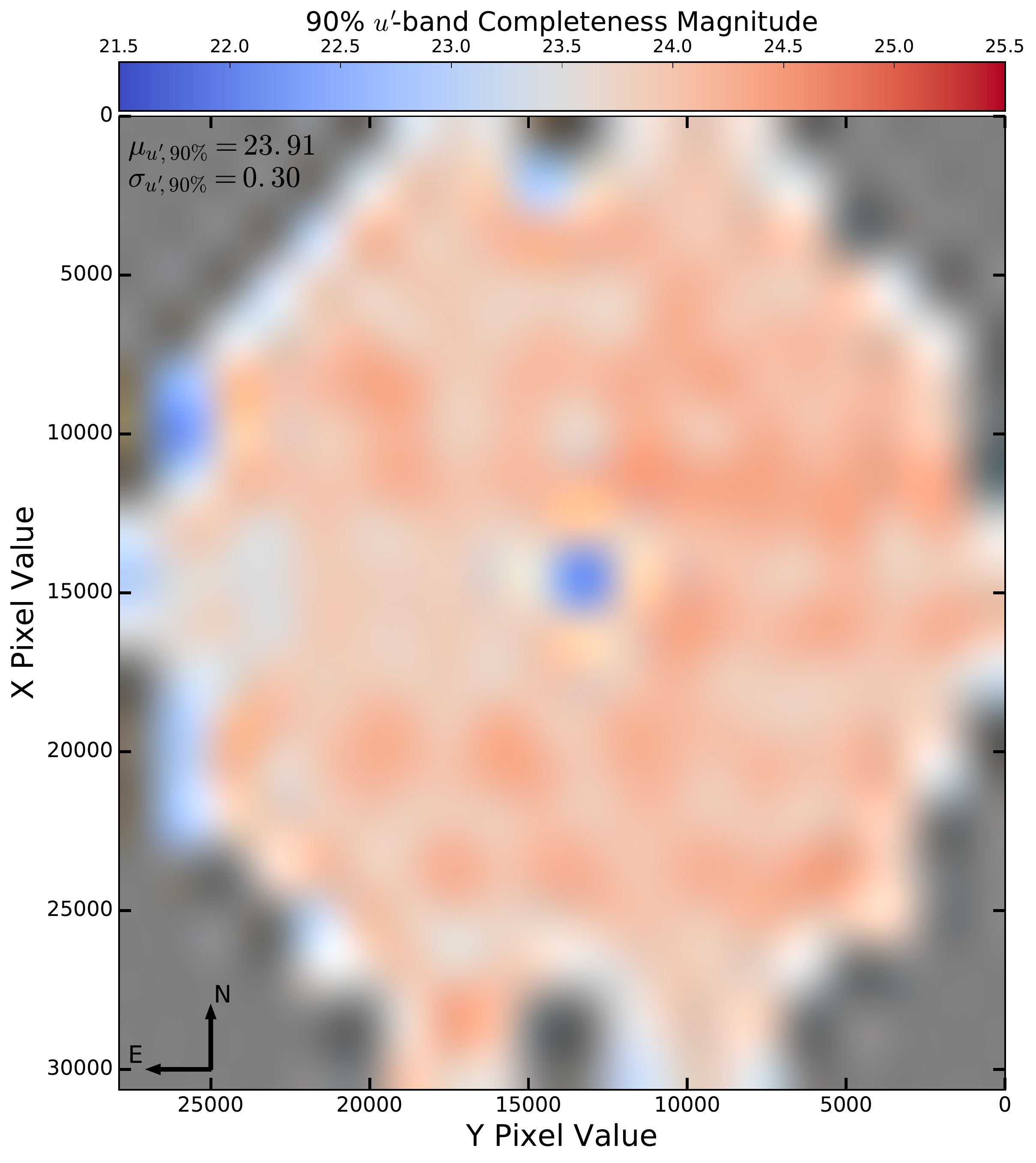}
\includegraphics[width=0.49\linewidth]{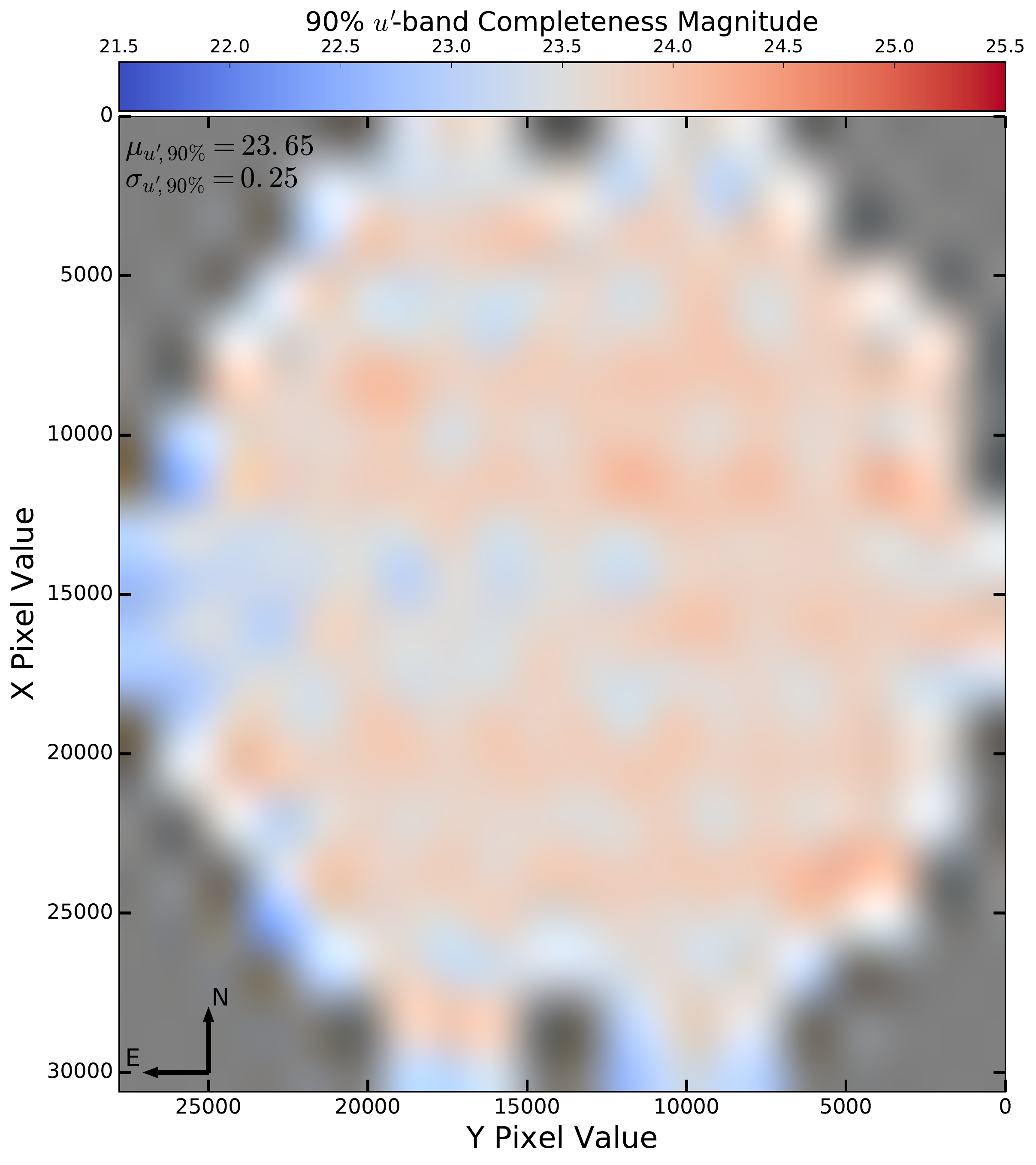}
\caption{Photometric depth variations in the $u'$-band based on artificial star experiments. Results for the central tile (Tile\,1) of SCABS are shown in the left column, and results representative of the Outer Ring (Tiles\,2--7) are shown on the right. Variations in the completeness magnitudes are parameterized by the colour bar, with the top row corresponding to the 50 percent completeness magnitude, and the bottom showing results for the 90 percent depth. Units along the axes are in DECam imaging pixels (see also Fig.\,\ref{fig:decam}), and the North and East cardinal directions are indicated in the lower-left corners of the panels. 1$\sigma$ variations are listed in the upper-left corners, below the mean photometric depths based on the variation maps upper-left corners, which are in good agreement with the adopted values shown in Tbl.\,\ref{tbl:complete} and Figs.\,\ref{fig:comptest_u}--\ref{fig:comptest_z}. Spurious drops in the depth variations represent regions that were under-sampled during the binning, and not considered to be physical.}
\label{fig:compmap_u}
\end{figure*}

\begin{figure*}
\centering
\includegraphics[width=0.49\linewidth]{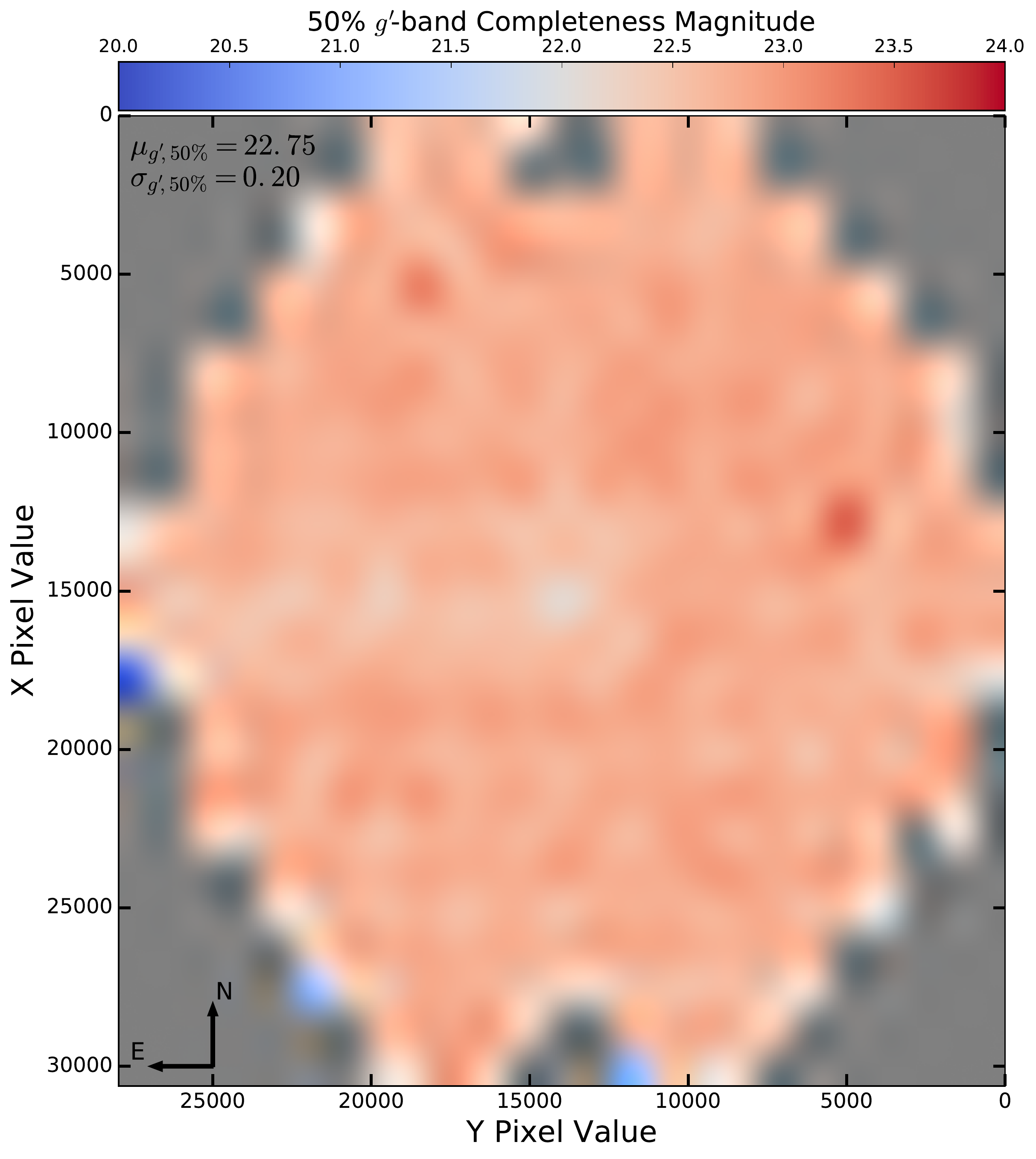}
\includegraphics[width=0.49\linewidth]{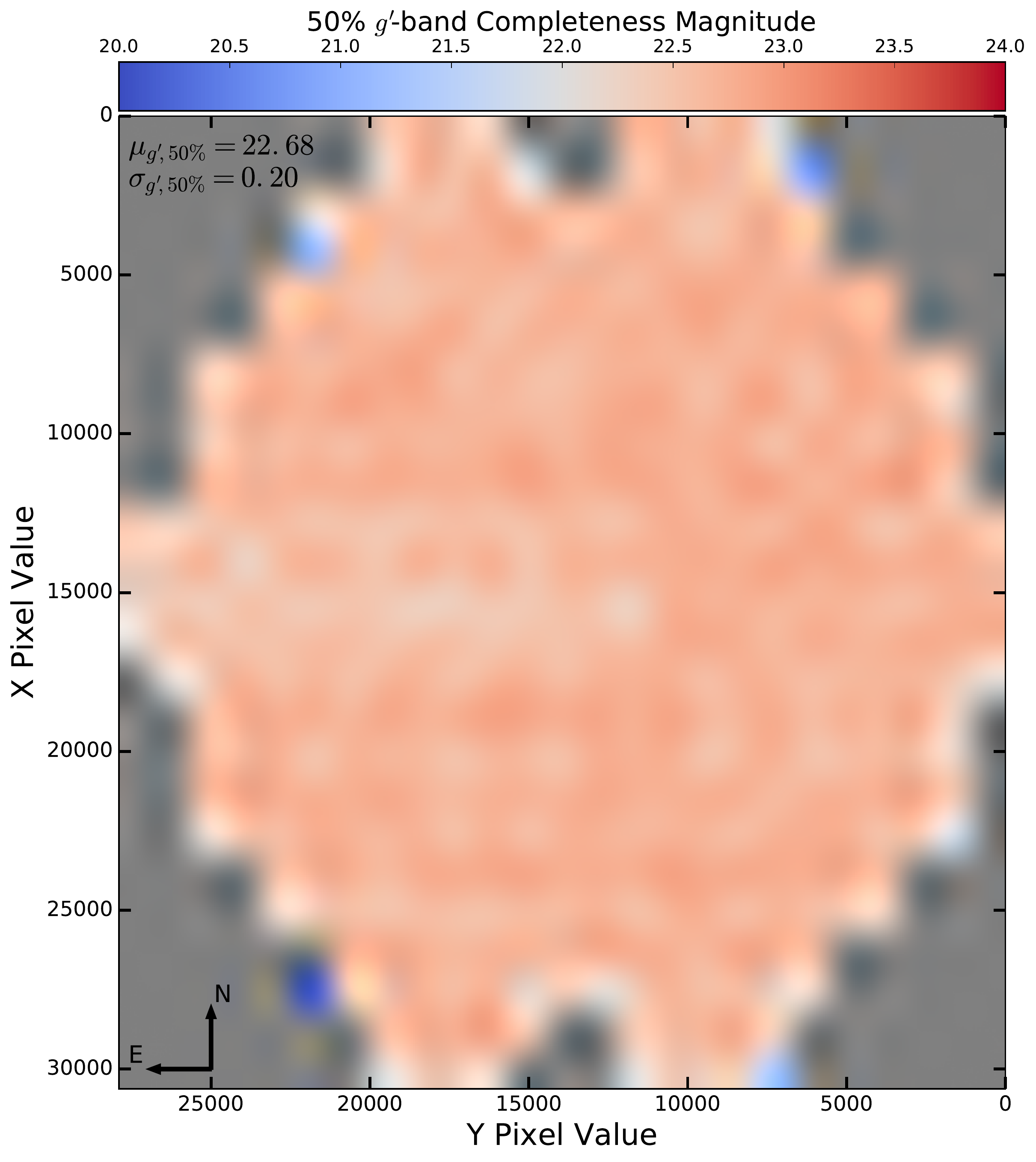}
\includegraphics[width=0.49\linewidth]{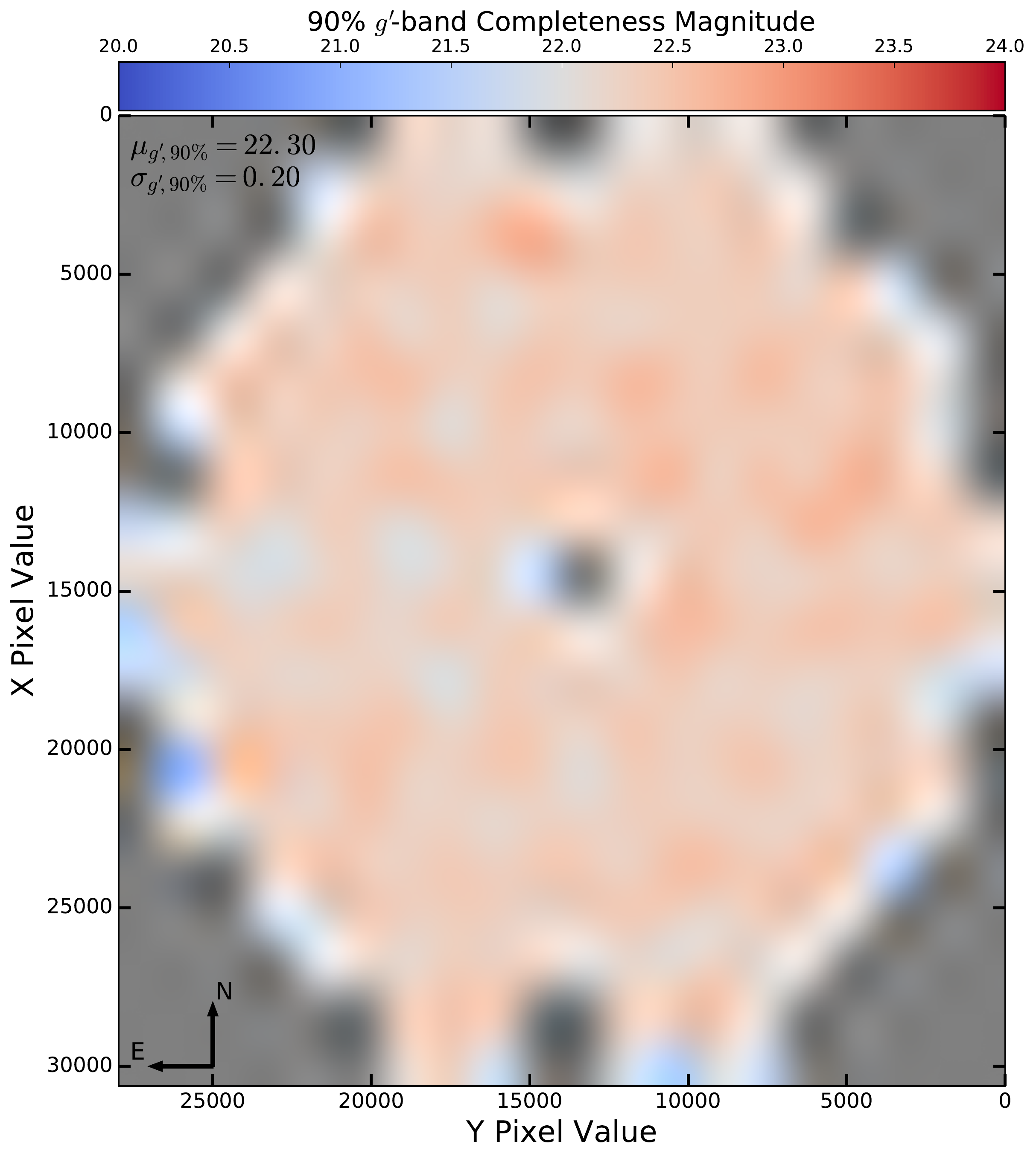}
\includegraphics[width=0.49\linewidth]{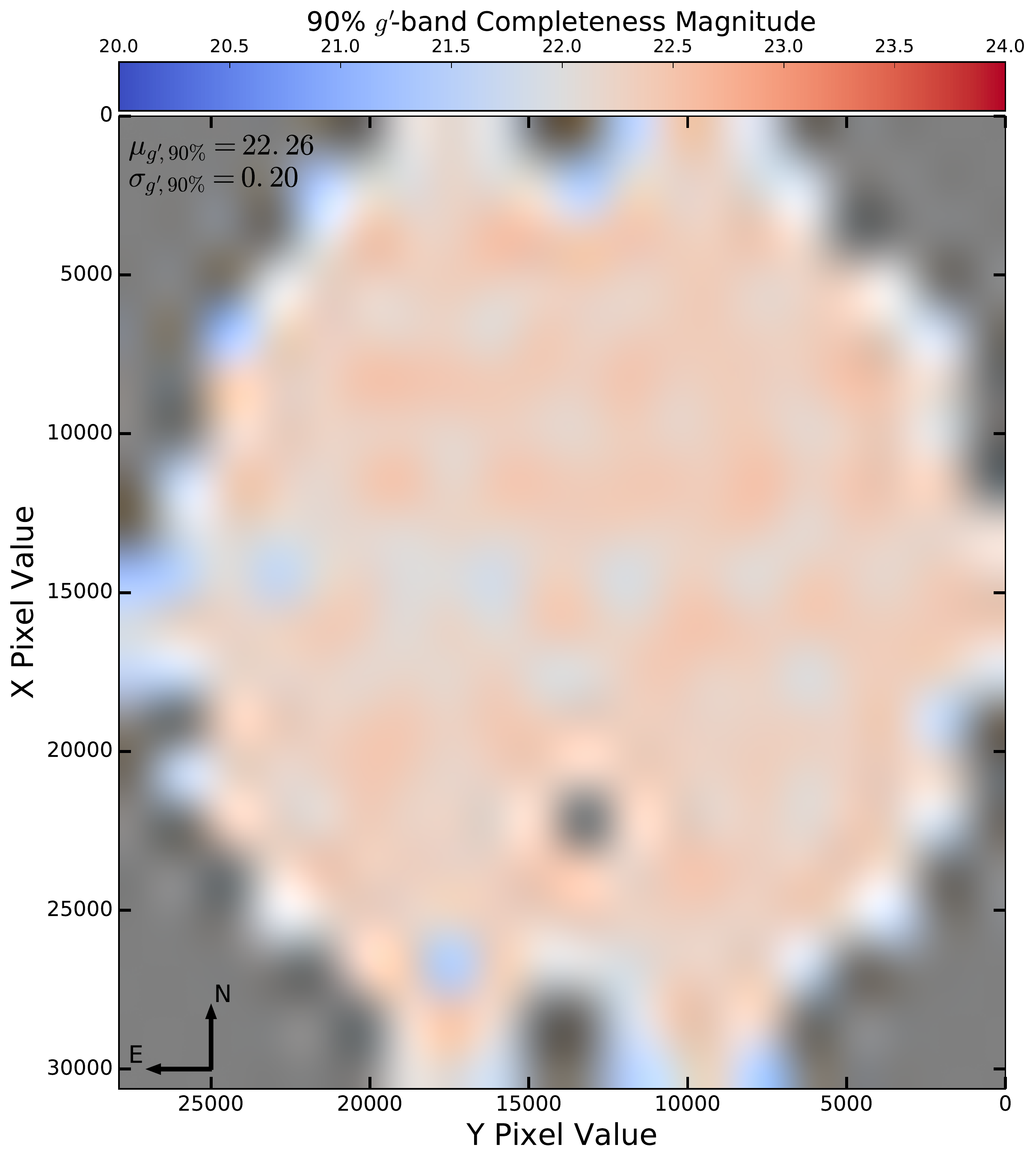}
\caption{Photometric depth variations in the $g'$-band based on artificial star experiments. See Fig.\,\ref{fig:compmap_u} for a detailed description of the Figure.}
\label{fig:compmap_g}
\end{figure*}

\begin{figure*}
\centering
\includegraphics[width=0.49\linewidth]{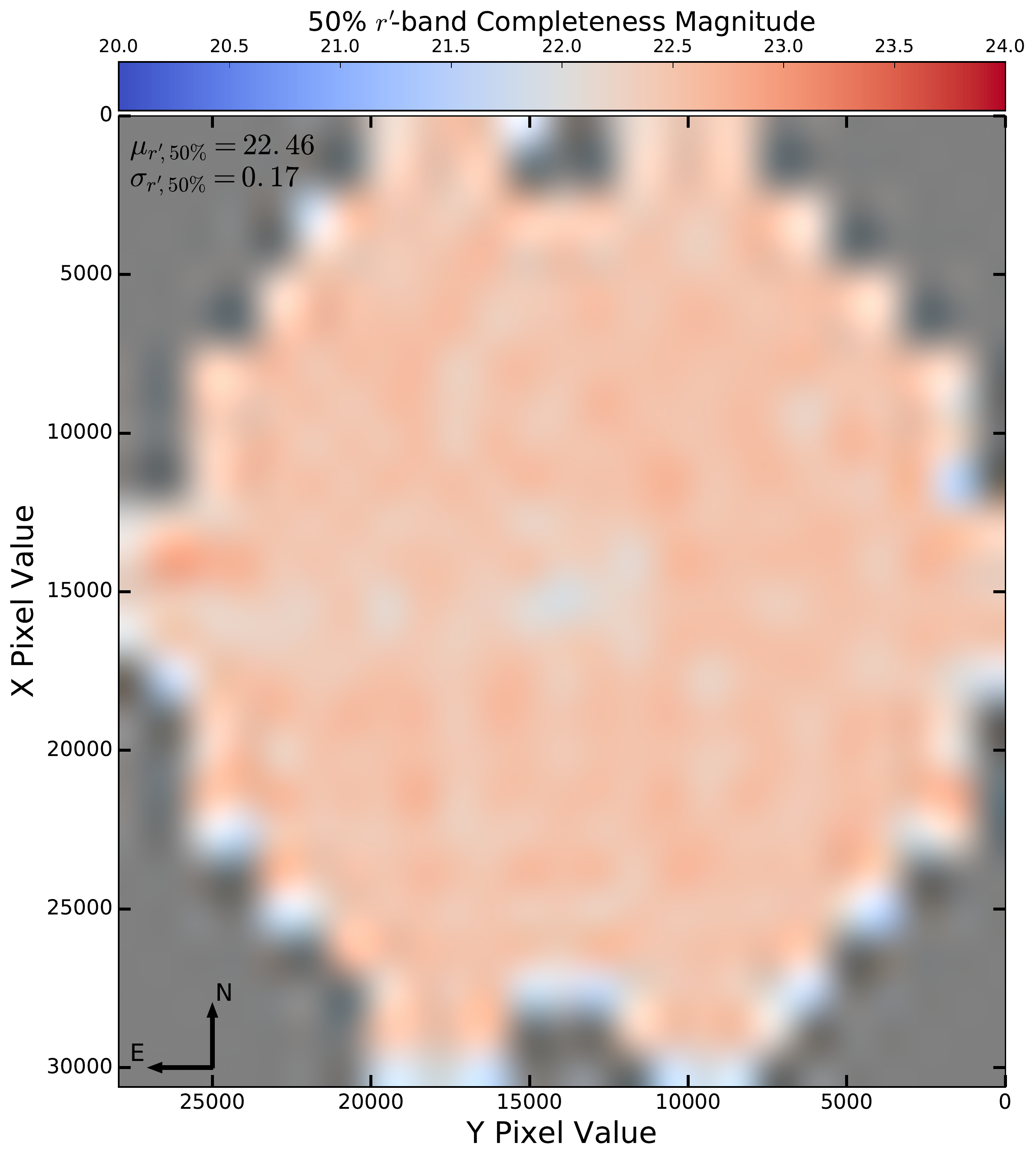}
\includegraphics[width=0.49\linewidth]{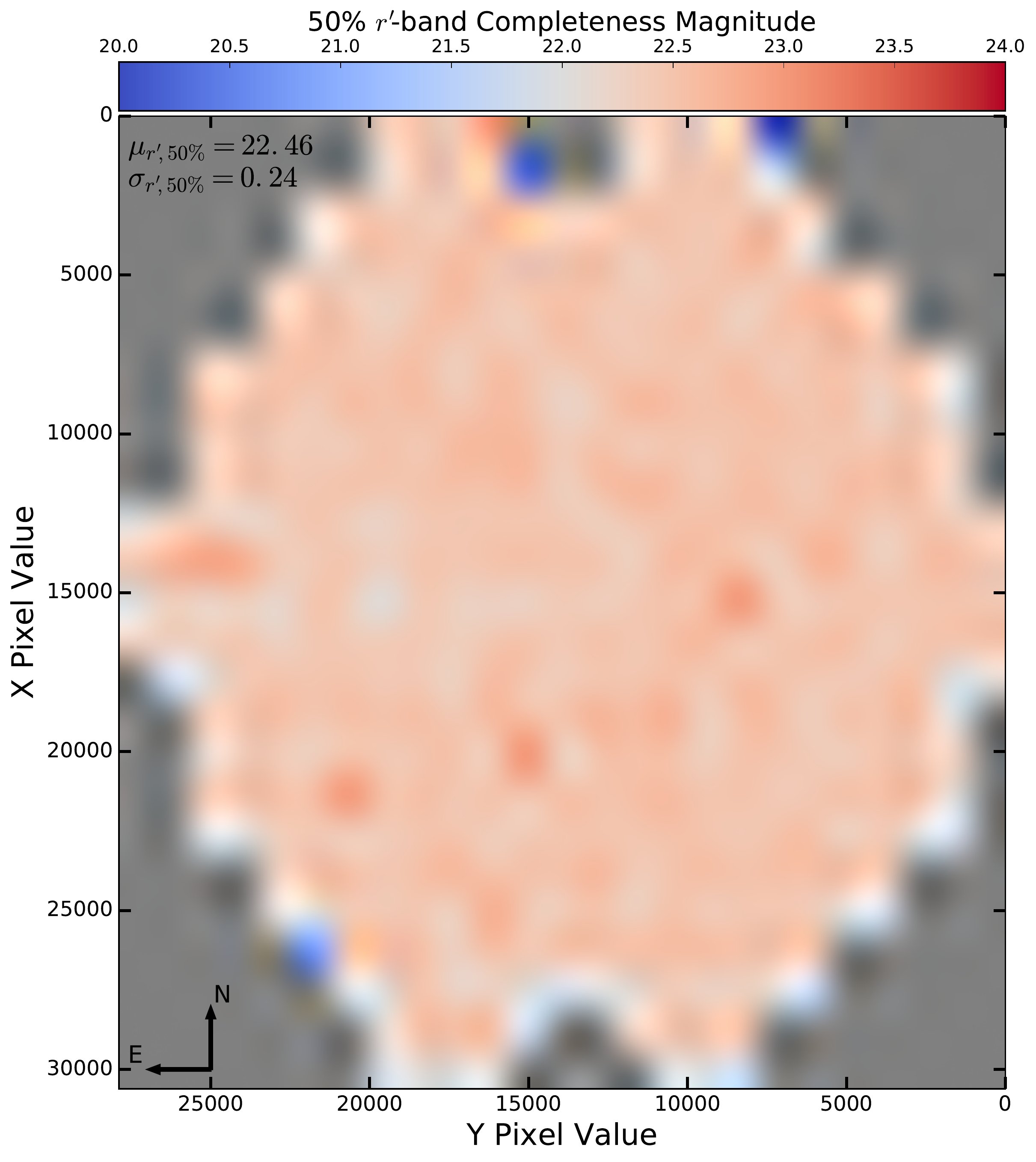}
\includegraphics[width=0.49\linewidth]{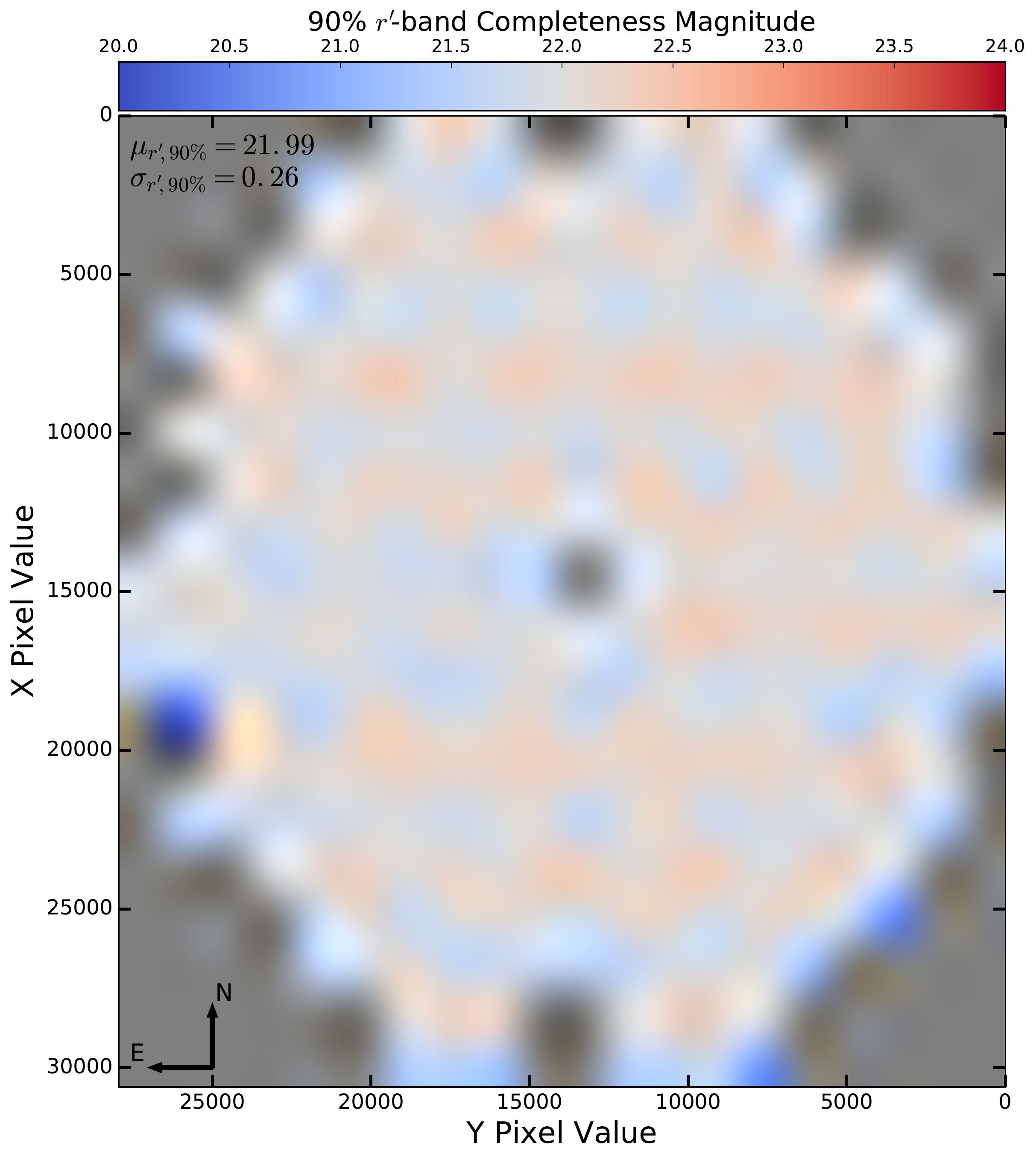}
\includegraphics[width=0.49\linewidth]{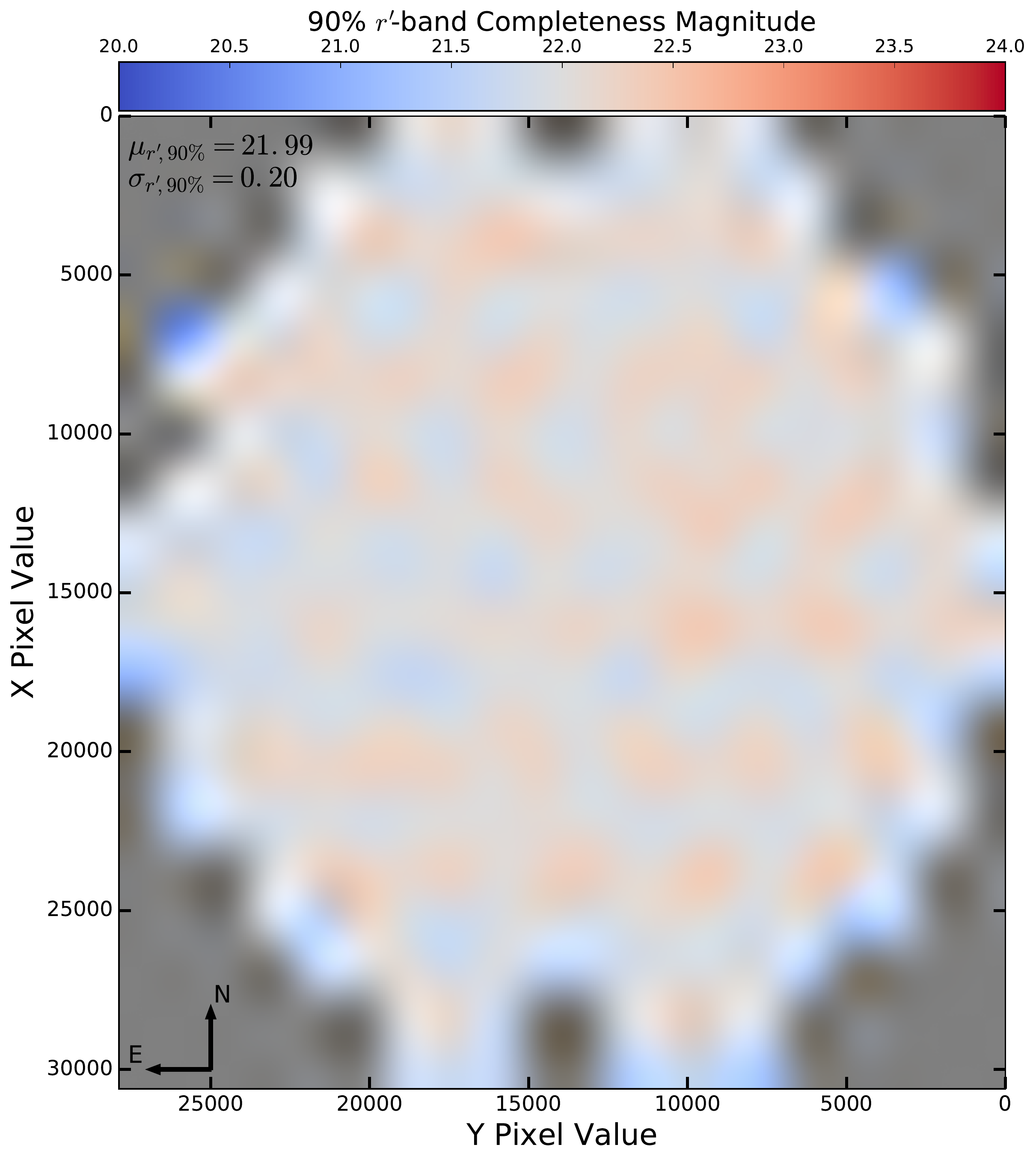}
\caption{Photometric depth variations in the $r'$-band based on artificial star experiments. See Fig.\,\ref{fig:compmap_u} for a detailed description of the Figure.}
\label{fig:compmap_r}
\end{figure*}

\begin{figure*}
\centering
\includegraphics[width=0.49\linewidth]{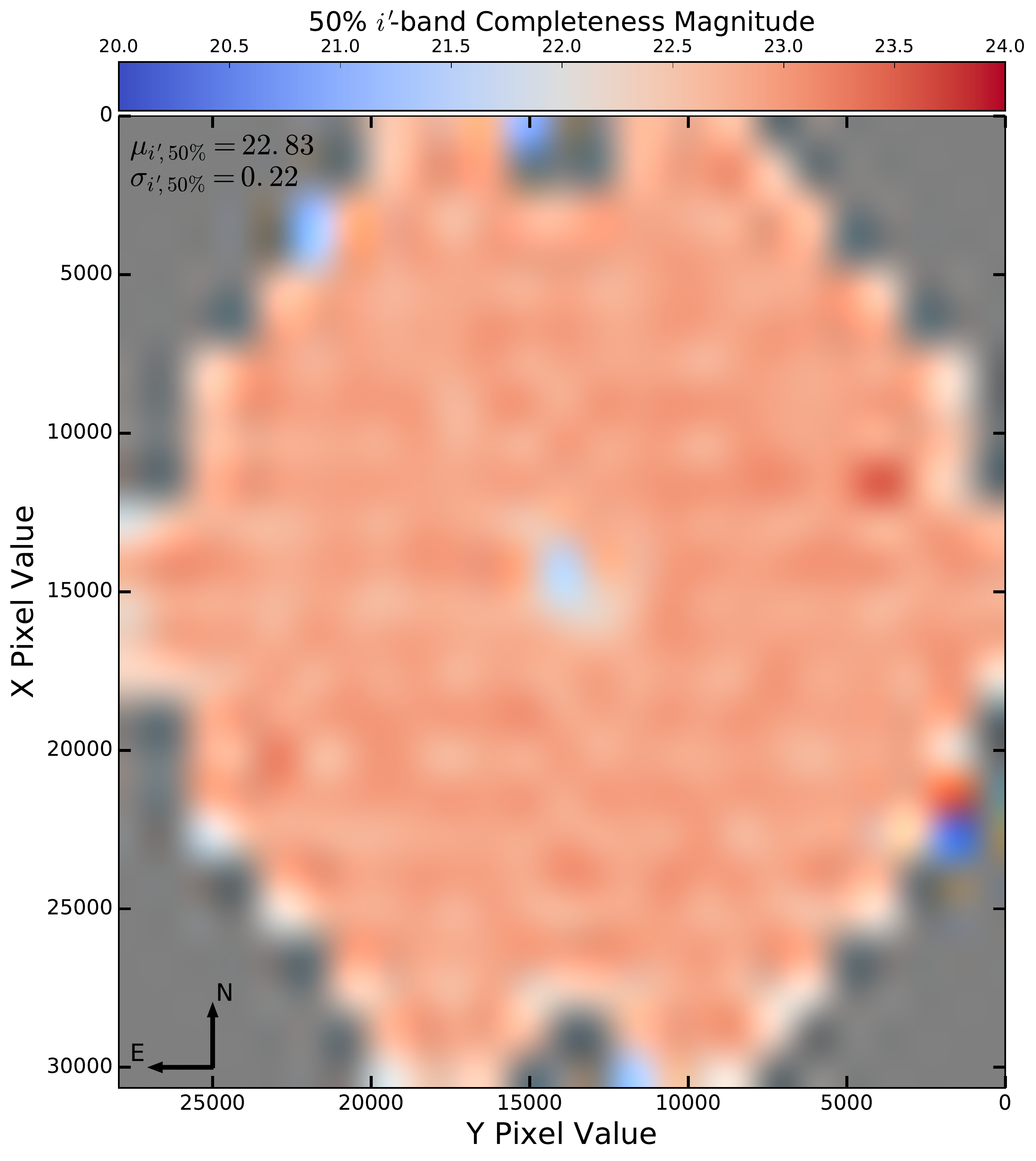}
\includegraphics[width=0.49\linewidth]{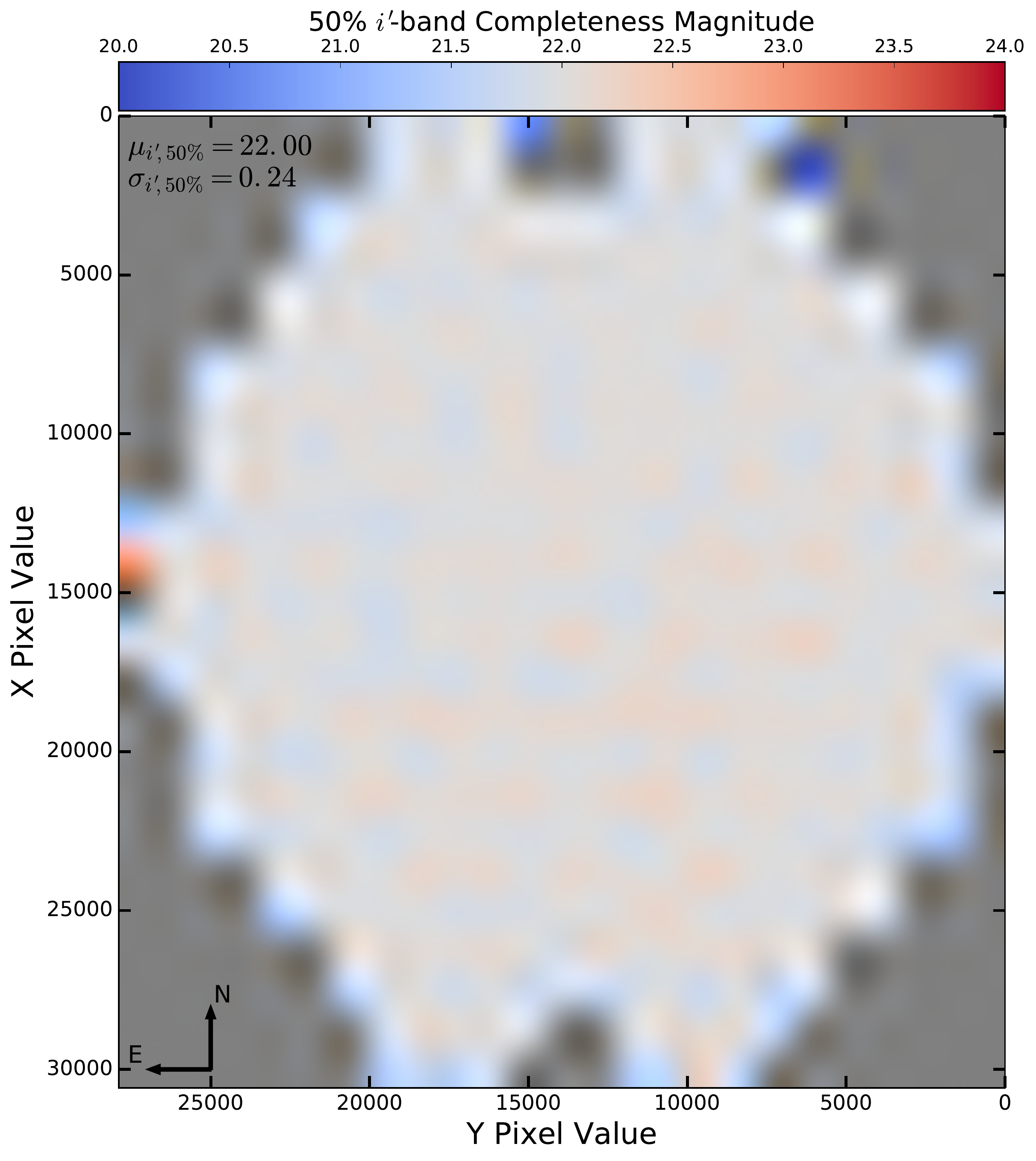}
\includegraphics[width=0.49\linewidth]{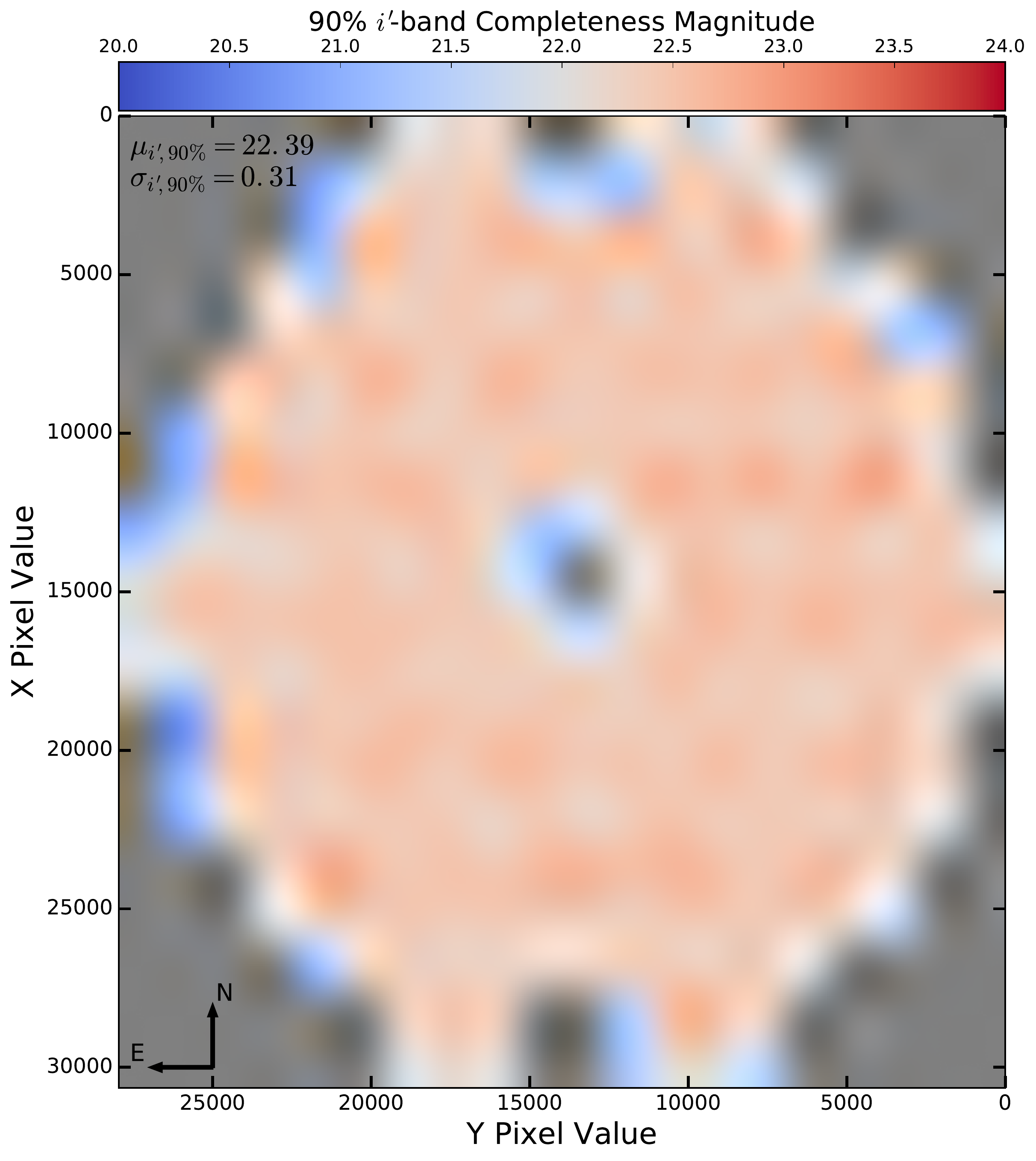}
\includegraphics[width=0.49\linewidth]{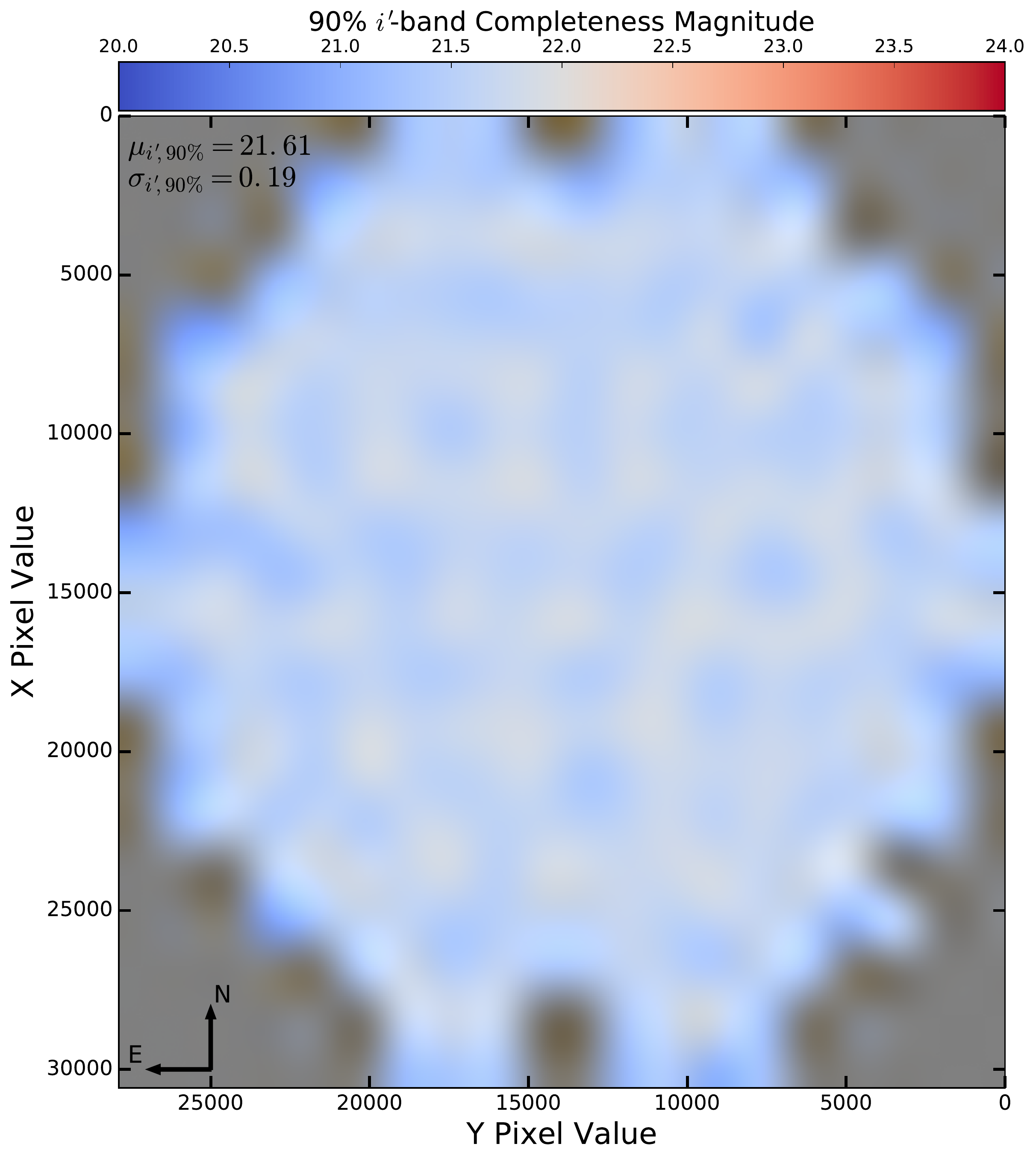}
\caption{Photometric depth variations in the $i'$-band based on artificial star experiments. See Fig.\,\ref{fig:compmap_u} for a detailed description of the Figure.}
\label{fig:compmap_i}
\end{figure*}

\begin{figure*}
\centering
\includegraphics[width=0.49\linewidth]{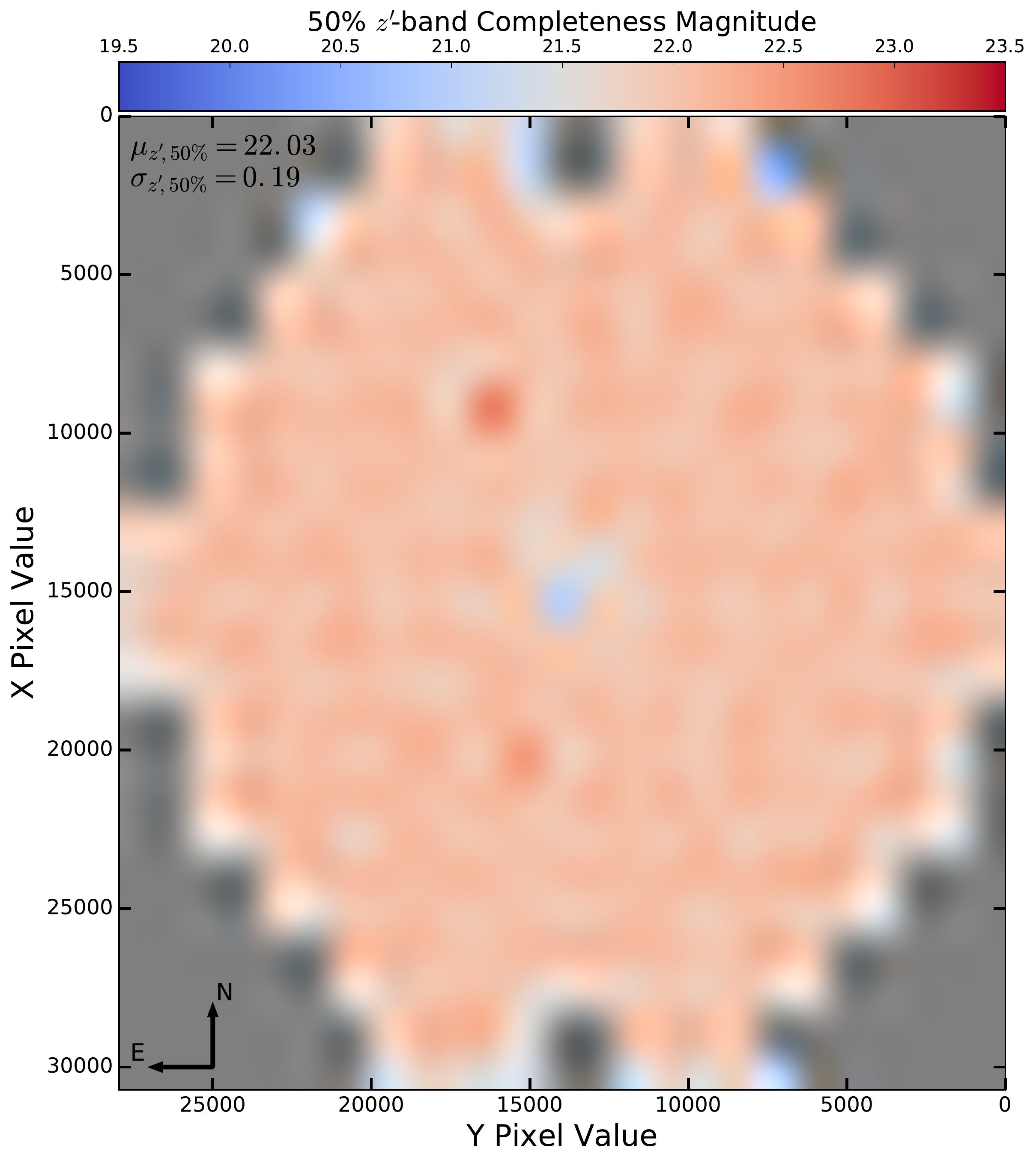}
\includegraphics[width=0.49\linewidth]{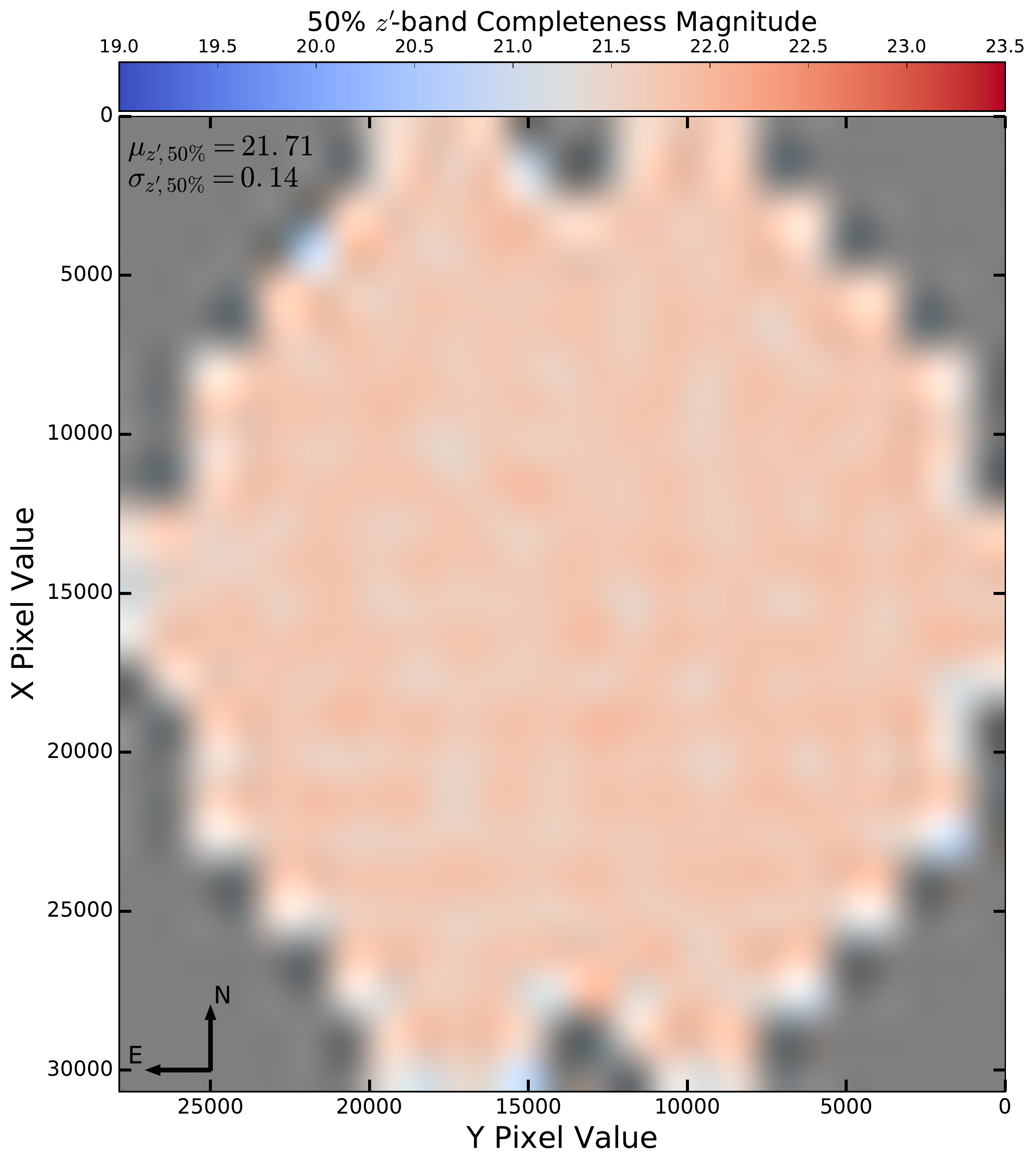}
\includegraphics[width=0.49\linewidth]{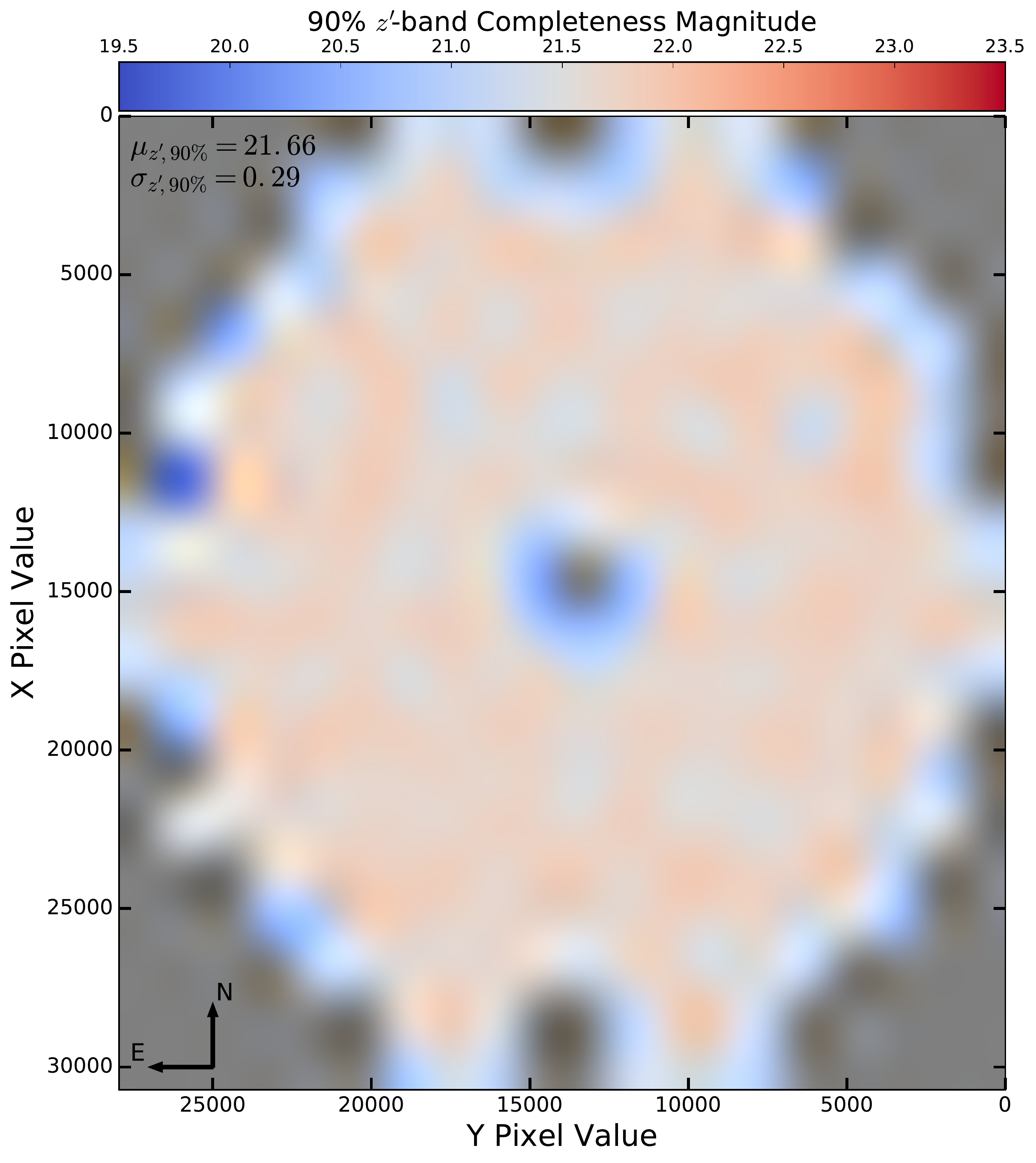}
\includegraphics[width=0.49\linewidth]{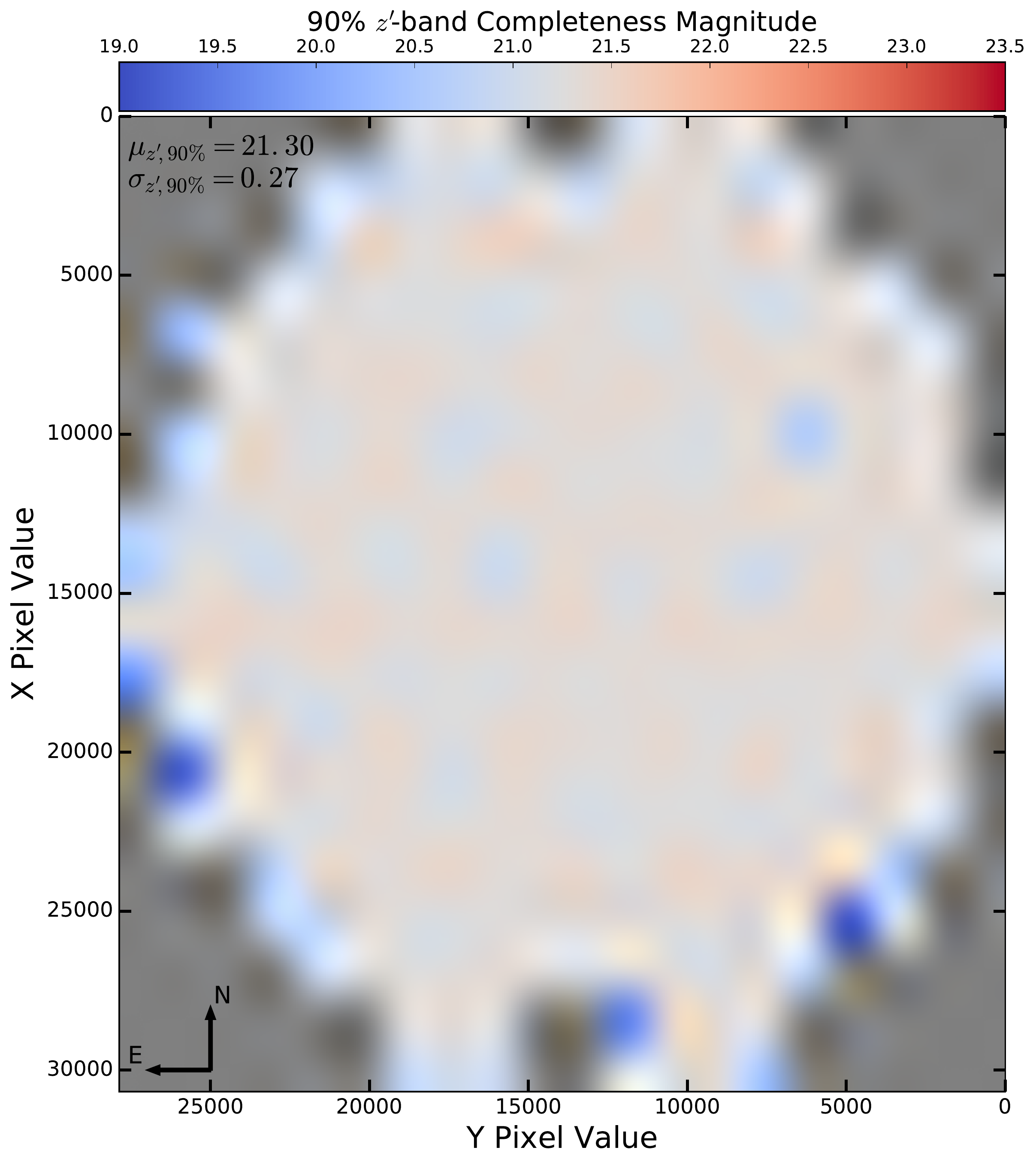}
\caption{Photometric depth variations in the $z'$-band based on artificial star experiments. See Fig.\,\ref{fig:compmap_u} for a detailed description of the Figure.}
\label{fig:compmap_z}
\end{figure*}

The results of the artificial star experiments are listed in Table\,\ref{tbl:complete}, and shown in Figs.\,\ref{fig:comptest_u}--\ref{fig:compmap_z}. The lower panels of Figs.\,\ref{fig:comptest_u}--\ref{fig:comptest_z} show, for each of the five filters in Tiles\,1 and 2, the fraction of recovered mock sources as a function of input magnitude. The blue shading bounded by dashed black lines indicates areas corresponding to 100, 90, and 50 percent completeness, with darker shading indicating higher completeness. The solid red relations indicate spline fits to the blue dashed curves, which represent the results of the experiments. In all cases, the analytic red curves represent the data well, and are used to infer the numerical completeness limits as labelled, and listed in Table\,\ref{tbl:complete}. The upper panels of Figs.\,\ref{fig:comptest_u}--\ref{fig:comptest_z} illustrate the robustness of the recovered photometry of the mock stars. The difference between the input PSF-based and recovered {\sc mag\_auto} magnitudes is shown as a function of input brightness, with a non-parametric linear regression and associated $1\sigma$ bootstrapped errors\footnote{Using the {\sc Python/AstroML} package.} shown by the red curves. These panels suggest that our point-source photometry is reliable within $\lesssim0.1$\,mag until at least the 90 percent completeness magnitude of a given band. Beyond this limit, in all bands, our recovered photometry predicts increasingly fainter sources relative to the input PSF model magnitudes, up to roughly our 50 percent completeness limits. These differences are due to the elliptical aperture used by {\sc SE}'s {\sc mag\_auto} not accounting for the tails of the PSF models at faint magnitudes, and do not exceed $\sim0.2$\,mag. Our listed PSF-based photometry accounts for this offset, which in any case, we conservatively include in our listed systematic error budgets.

Figs.\,\ref{fig:compmap_u}--\ref{fig:compmap_z} show the photometric stability across the field of view of DECam. The left-hand panels show results for Tile\,1, with Tiles\,2--7 shown on the right, while the upper and lower rows correspond to the 50 and 90 percent completeness limits, respectively. In image pixel space, the completeness test images are sliced into 2D bins, and 50 and 90 percent completeness limits are calculated for individual bins. The resulting grid of limiting magnitudes is then used to infer values across the entire image using {\sc lanczos} interpolation, and is indicated by the colour maps with sensitivity increasing from blue to red.

Small patches that sharply transition to shallower (or NaN) limits persist across several of the panels of Figs.\,\ref{fig:compmap_u}--\ref{fig:compmap_z}. These artifacts are a result of the stochastic process of adding mock stars in regions that contain extended background sources, combined with the random sampling of magnitudes assigned to the mock stellar catalogues. Testing larger and/or smaller bin sizes results in these patches shifting position among the various panels, and thus we do not consider them to be physical in nature. With this in mind, the median limiting magnitudes across the field of view are shown in the upper left corner of each panel, which agree well with the values based on Figs.\,\ref{fig:comptest_u}--\ref{fig:comptest_z} and listed in Table\,\ref{tbl:complete}.

The artificial star experiments indicate that for the Tiles\,2--7, we reach 90 percent completeness magnitudes of 23.61, 22.26, 22.02, 21.63, and 21.33\,mag in the $u'$, $g'$, $r'$, $i'$, and $z'$ filters, respectively, with correspondingly fainter 50 percent completeness limits of 24.08, 22.67, 22.46, 22.00, and 21.71\,mag. Meanwhile, the deeper Tile\,1 $u'$-, $i'$-, and $z'$-band imaging is reflected by the $\sim0.3$, $\sim0.8$, and $\sim0.3\,{\rm mag}$ deeper 50 percent and 90 percent completeness limits, respectively. We also investigate the areal stability of the photometric sensitivities, with completeness estimate dispersions based on the four faintest magnitude bins listed alongside the derived magnitude limits in Table\,\ref{tbl:complete}. We find that the photometric sensitivities are very stable with variations of, at worst, $\lesssim0.3\,{\rm mag}$ across the $\sim3\,{\rm deg}^2$ DECam footprint, and generally tend towards variations of $\lesssim0.2\,{\rm mag}$. Encouragingly, the adoption of the Tile\,2 limits for Tiles\,3--7 appears reasonable, as there is a near perfect agreement for the $r'$-band results, and a $\lesssim0.1\,{\rm mag}$ discrepancy for the $g'$-band, which is likely due to seeing variations during the observations and is in any case, less than the photometric variation across the DECam field of view.

%%%%%%%%%%%%%%%%%%%%%%%%%%%%%%%
%%%%%%%%%%%%%%%%%%%%%%%%%%%%%%%
%%%%%%%%%%%%%%%%%%%%%%%%%%%%%%%

\section{Source Catalogues}
\label{sec:scat}
The source catalogues are publicly available by querying \url{http://vizier.u-strasbg.fr/viz-bin/VizieR}. In accords with Table\,\ref{tbl:sources}, we provide photometric measurements for as few as $595\,136$ sources in the $u'$-band and up to $1\,490\,519$ sources with $i'$-band photometry. Table\,\ref{tbl:source_cats} summarizes the contents of the available measurements, alongside the corresponding catalogue parameter names (based on {\sc SE} where possible). A total of nine photometric measurements are provided with associated {\sc SE} error estimates, corresponding to fluxes within seven fixed-circular apertures, an elliptical aperture, and PSF-based magnitudes. Following the photometric measurements we list the adopted statistical and systematic error budgets (see \S\,\ref{sec:phot}). Several morphological parameters are also provided, including the {\sc SE} half-light radius, source ellipticity, isophotal area measured within five isocontours, and a compactness parameter. {\sc SE} measures isocontours at seven surface brightness levels, the areas within each are given by the {\sc iso\#} parameters in units of pix$^2$. Here we list five corresponding to the two faintest isocontours ({\sc iso$0,1$}) increasing to the smallest isophotal area containing the brightest pixels of the source ({\sc iso$7$}). Finally, for convenience, the foreground reddening in the corresponding band is listed based on the reddening maps of \cite{sch11}.
 
\begin{table}
	\centering
	\caption{Point source catalogue information with all magnitudes listed in AB.}
	\label{tbl:source_cats}
	\begin{tabular}{lcr}
		\hline
		Description 				&	Unit			& 	{\sc Parameter}		\\
		\hline
		\hline
		Description 				&	Unit			& 	{\sc Parameter}		\\
		\hline
		Right Ascension			&	deg.		&	{\sc alpha\_j2000}		\\
		Declination				&	deg.		&	{\sc delta\_j2000}		\\
		Fixed Aperture Magnitudes (7)	&	mag. 	&	{\sc mag\_aper}			\\
		Errors (7)					&	mag. 	&	{\sc magerr\_aper}		\\
		Elliptical Aperture Magnitude	&	mag.		&	{\sc mag\_auto}			\\
		Error						&	mag.		&	{\sc magerr\_auto}		\\
		PSF Magnitude				&	mag.		&	{\sc mag\_psf}			\\
		Error						&	mag.		&	{\sc magerr\_psf}		\\
		Statistical Error (see \S\,\ref{sec:phot})	&	mag.		&	{\sc stat\_err}	\\
		Systematic Error (see \S\,\ref{sec:phot})	&	mag.		&	{\sc sys\_err}	\\
		Effective Radius			&	pixel		&	{\sc flux\_radius}		\\
		Image FWHM				&	pixel		&	{\sc fwhm\_image}		\\
		World FWHM				&	deg.		&	{\sc fwhm\_world}		\\
		Major Axis				&	pixel		&	{\sc a\_image}			\\
		Minor Axis				&	pixel		&	{\sc b\_image}			\\
		Ellipticity					&	$1-\frac{b}{a}$	&	{\sc ellipticity}			\\
		Position Angle				&	deg.		&	{\sc theta\_image}		\\
		Isophotal Area 0			&	pixel$^2$	&	{\sc iso0}				\\
		Isophotal Area 1			&	pixel$^2$	&	{\sc iso1}				\\
		Isophotal Area 3			&	pixel$^2$	&	{\sc iso3}				\\
		Isophotal Area 5			&	pixel$^2$	&	{\sc iso5}				\\
		Isophotal Area 7			&	pixel$^2$	&	{\sc iso7}				\\
		Spread Parameter			&			&	{\sc spread\_model}		\\
		Spread Error				&			&	{\sc spreaderr\_model}	\\
		Phototmetry Flag			&			&	{\sc FLAGS}			\\
		Foreground Reddening		&	mag.		&	{\sc reddening}			\\
		\hline
	\end{tabular}
\end{table}

%%%%%%%%%%%%%%%%%%%%%%%%%%%%%%%
%%%%%%%%%%%%%%%%%%%%%%%%%%%%%%%
%%%%%%%%%%%%%%%%%%%%%%%%%%%%%%%

\section{Discussion}
\label{sec:disc}
The value of these data covers myriad potential lines of investigation. For example, the 90 percent completeness depths of these data are sufficient to detect $\gtrsim95$ percent of globular clusters (GCs) in the region, assuming a GC luminosity function peak at $m_V\approx-7.5$\,mag \citep[e.g.][]{har01} and dispersion of $\sim1.0$\,mag \citep{vil10}. Reaching out to $\sim140$\,kpc from \cena, the suite of colour indices are sufficient to distinguish most GCs from both foreground stellar sources, and point-like background sources that morphologically masquerade as GCs. Moreover, the extent of SCABS reaches into the extreme halo of \cena\ and thus should begin to probe GCs associated with the intra-group medium of Centaurus A, including any dwarfs in the region. Altogether, the potential for these data to identify the vast majority of GCs associated with \cena\ represents a powerful method of constraining its mass assembly history through cosmic time.

The serendipitous location of \cena\ at relatively low Galactic latitude ($b=19.42^\circ$) makes this dataset valuable for more than strictly extragalactic investigation. The wide-field coverage of DECam, with pointings mildly off of the Galactic plane makes the removal of foreground stars challenging for background science, but the sampling of the full optical SED eases this task and thus can provide a rich list of foreground star photometry. Indeed, the $\sim21\,{\rm deg}^2$ of coverage samples foreground stars through the thin and thick Galactic disks, and out into the halo toward \cena, potentially providing a rich catalogue of foreground stars with ten permutations of optical colour indices. By comparison to state-of-the-art stellar atmospheric models, rich ancillary science including studies of the ages and metallicities of the various populations of Galactic foreground stars are possible.

The doubling of the Local Group dwarf galaxy population in the past decade \citep[e.g.][]{wil05,bel06,bel07,mcc09,mcc12} provides a potent window on the epoch of the first galaxies via near-field cosmology studies coupled with simulations \citep[e.g.][]{sal09,bov09,bov11a,bov11b}; however, little is known about dwarf galaxy populations beyond the Local Group. To this end, rich populations of dwarf galaxies are being discovered in nearby galaxy clusters like Virgo and Fornax \citep{mun15,san16} and \cena\ itself \citep{kar07,crn14,crn15,mul15,mul16} which provide direct observational tests of the bottom-up hierarchical formation of their hosts as favoured by $\Lambda$\,Cold Dark Matter cosmology \citep[e.g.][]{kly99,moo99}. These observations, in conjunction with the identification of the group's extended GC system, will then be of excellent utility in placing new constraints on the assembly history of this iconic galaxy and its group environment, which is the natural extension of the ongoing detailed Local Volume studies, such as SDSS, PanSTARRS, PANDAS, and the {\it Dark Energy Survey} \citep[e.g.][]{yor00,abb06,mcc09,kai10}.

While the depth is not sufficient to probe the high-redshift ($z\!\gtrsim\!1$) universe, the optical luminosity function (LF) of intermediate redshift galaxies is sensitive to the star formation/morphological properties of the underlying galaxy population. Meanwhile, similarly deep near-infrared (NIR) observations have already been conducted and will extend the SED coverage yet further. Together with the optical data discussed here these data will probe the mass function (MF) of background cluster galaxies \citep[e.g.][]{mad98}. Since the evolution of the MF is directly predicted from hierarchical galaxy formation models that incorporate theoretical SEDs \citep[e.g.\ {\sc galform}][]{col00}, the full NUV-NIR LFs represent an excellent test for model predictions such as these. Altogether the depth of the SCABS data provide an excellent opportunity to derive galaxy parameters for a rich sample of galaxies out to $z\approx1$.

%%%%%%%%%%%%%%%%%%%%%%%%%%%%%%%
%%%%%%%%%%%%%%%%%%%%%%%%%%%%%%%
%%%%%%%%%%%%%%%%%%%%%%%%%%%%%%%

\section{Summary}
This paper presents new optical ($u'g'r'i'z'$) observations of the central $\sim21\,{\rm deg}^2$ ($\sim62\,000\,{\rm kpc}^2$) region of the Centaurus A galaxy group centred on NGC\,5182, as part of the {\it Survey of Centaurus A's Baryonic Structures (SCABS)}. The observations have a raw data reduction conducted by the {\it CTIO-DECam Community Pipeline} \citep[][v.3.1.1]{val14}, from which we derive photometric and astrometric calibrations using our custom built post-processing pipeline based on the {\sc Astromatic} software suite \citep{ber96,ber02,ber06,ber11}. Individual frames are aligned to a common world coordinate solution, and co-added to produce images which are sufficient for the analysis of point-like, or mildly extended sources. Artificial star experiments are conducted to derive 50 and 90 percent point-source completeness estimates, finding 90 percent completeness magnitudes of at least $23.62$, $22.27$, $22.00$, $21.63$, and $21.34$\,AB mag in the $u'$-, $g'$-, $r'$-, $i'$-, and $z'$-bands, respectively, with very stable photometric sensitivity across the field.

We release our source catalogues for public use, which can be used to, as non-exhaustive examples, probe the compact stellar systems (UCDs and GCs) of the Centaurus A galaxy group in the context of their stellar population parameters and \cena's mass assembly history, study the background universe out to $z\approx1.0$, and probe the properties of the distinct populations of Galactic foreground stars in the direction of \cena. We look forward to releasing future source catalogues including the central $\sim72\,{\rm deg}^2$ around Centaurus A, as well as finishing a detailed background subtraction that will yield multi-band photometry of \cena's rich dwarf galaxy population, both old and new. Furthermore, deep NIR imaging is already in-hand, which will be added to the current imaging to provide full NUV-NIR SEDs for nearly one million sources in the field and provide myriad science results in the upcoming years.

%%%%%%%%%%%%%%%%%%%%%%%%%%%%%%%
%%%%%%%%%%%%%%%%%%%%%%%%%%%%%%%
%%%%%%%%%%%%%%%%%%%%%%%%%%%%%%%

\section*{Acknowledgements}
We wish to thank Simon \'Angel, Yasna Ordenes-Brice\~no, Mirko Simunovic, and Hongxin Zhang for fruitful discussion, and especially Eric Peng for additionally providing us with catalogues of new confirmed foreground stars and GCs prior to publication. M.A.T.\ acknowledges the financial support through an excellence grant from the ``Vicerrector\'ia de Investigaci\'on" and the Institute of Astrophysics Graduate School Fund at Pontificia Universidad Cat\'olica de Chile and the European Southern Observatory Graduate Student Fellowship program. T.H.P. acknowledges support by a FONDECYT Regular Project Grants (No.~1121005 and No.~1161817) and the BASAL Center for Astrophysics and Associated Technologies (PFB-06). M.S.B. was supported in part by FONDECYT Project Grant (No. 3130549).\\
This research has made use of the NASA Astrophysics Data System Bibliographic Services, the NASA Extragalactic Database, and the SIMBAD database, operated at CDS, Strasbourg, France \citep{wen00}. Software used in the analysis includes the {\sc Python/NumPy} v.1.11.0 and {\sc Python/Scipy} v0.17.0 \citep[][\url{http://www.scipy.org/}]{jon01,van11}, {\sc Python/astropy} \citep[v1.1.1;][\url{http://www.astropy.org/}]{ast13}, {\sc Python/matplotlib} \citep[v1.5.1;][\url{http://matplotlib.org/}]{hun07}, {\sc Python/scikit-learn} \citep[v0.16.1;][\url{http://scikit-learn.org/stable/}]{ped12}, and {\sc Python/astroML} \citep[v0.3;][\url{http://www.astroml.org/}]{van12} packages.\\
This work is based on observations at Cerro Tololo Inter-American Observatory, National Optical Astronomy Observatory (CNTAC Prop. ID: 2014A-0610; PI: Matthew Taylor), which is operated by the Association of Universities for Research in Astronomy (AURA) under a cooperative agreement with the National Science Foundation.
This project used data obtained with the Dark Energy Camera (DECam), which was constructed by the Dark Energy Survey (DES) collaboration. Funding for the DES Projects has been provided by the U.S.\ Department of Energy, the U.S.\ National Science Foundation, the Ministry of Science and Education of Spain, the Science and Technology Facilities Council of the United Kingdom, the Higher Education Funding Council for England, the National Center for Supercomputing Applications at the University of Illinois at Urbana-Champaign, the Kavli Institute of Cosmological Physics at the University of Chicago, Center for Cosmology and Astro-Particle Physics at the Ohio State University, the Mitchell Institute for Fundamental Physics and Astronomy at Texas A\&M University, Financiadora de Estudos e Projetos, Funda\c{c}\~ao Carlos Chagas Filho de Amparo, Financiadora de Estudos e Projetos, Funda\c{c}\~ao Carlos Chagas Filho de Amparo \'a Pesquisa do Estado do Rio de Janeiro, Conselho Nacional de Desenvolvimento Cient\'ifico e Tecnol\'ogico and the Minist\'erio da Ci\^{e}ncia, Tecnologia e Inova\c{c}\~ao, the Deutsche Forschungsgemeinschaft and the Collaborating Institutions in the Dark Energy Survey. The Collaborating Institutions are Argonne National Laboratory, the University of California at Santa Cruz, the University of Cambridge, Centro de Investigaciones En\'ergeticas, Medioambientales y Tecnol\'ogicas--Madrid, the University of Chicago, University College London, the DES-Brazil Consortium, the University of Edinburgh, the Eidgen\~ossische Technische Hochschule (ETH) Z\~urich, Fermi National Accelerator Laboratory, the University of Illinois at Urbana-Champaign, the Institut de Ci\`encies de l'Espai (IEEC/CSIC), the Institut de F\'sica d'Altes Energies, Lawrence Berkeley National Laboratory, the Ludwig-Maximilians Universit\"at M\~unchen and the associated Excellence Cluster Universe, the University of Michigan, the National Optical Astronomy Observatory, the University of Nottingham, the Ohio State University, the University of Pennsylvania, the University of Portsmouth, SLAC National Accelerator Laboratory, Stanford University, the University of Sussex, and Texas A\&M University.
%
%%%%%%%%%%%%%%%%%%%%%%%%%%%%%%%%%%%%%%%%%%%%%%%%%%%
%
%%%%%%%%%%%%%%%%%%%%% REFERENCES %%%%%%%%%%%%%%%%%%
%
%% The best way to enter references is to use BibTeX:
%
%%\bibliographystyle{mnras}
%%\bibliography{example} % if your bibtex file is called example.bib
%
%
%% Alternatively you could enter them by hand, like this:
%% This method is tedious and prone to error if you have lots of references
\setlength\parskip{0.01\baselineskip}

%
%%%%%%%%%%%%%%%%%%%%%%%%%%%%%%%%%%%%%%%%%%%%%%%%%%%
%
%%%%%%%%%%%%%%%%%% APPENDICES %%%%%%%%%%%%%%%%%%%%%
%
%%\appendix
%%
%%\section{Some extra material}
%%
%%If you want to present additional material which would interrupt the flow of the main paper,
%%it can be placed in an Appendix which appears after the list of references.
%
%%%%%%%%%%%%%%%%%%%%%%%%%%%%%%%%%%%%%%%%%%%%%%%%%%%
%
%
%% Don't change these lines
\bsp	% typesetting comment
\label{lastpage}
\end{document}